\documentclass[galley,usenatbib]{mn2e}
\usepackage{graphicx}
\usepackage[figuresright]{rotating}
\usepackage{subfigure}
\usepackage{longtable}
\usepackage{txfonts}
\newcommand{\ii}{\'\i}
\newcommand\ion[2]{#1$\,${\scshape{#2}}}
\newcommand\ohlog{\rm \log(O/H)}
                                             
\title[A study of the interacting system AM 2306-721]
{Kinematics and physical properties of Southern interacting galaxies:
the minor merger AM 2306-721}

\author[Krabbe et al.]
{A.~C.~Krabbe$^{1,2}$\thanks{E-mail:angela.krabbe@ufrgs.br}, 
M.~G.~Pastoriza$^1$, Cl\'audia~Winge$^3$, I.~Rodrigues$^4$, and D.~L.~Ferreiro$^5$
\\
$^1$ Instituto de F\ii sica, Universidade Federal 
do Rio Grande do Sul, Av.~Bento Gon\c{c}alves, 9500, 
Cep 91359-050, Porto Alegre, RS, Brazil\\
$^2$ Southern Astrophysical Research Telescope, c/o AURA Inc., Casilla 603, La Serena, Chile\\
$^3$ Gemini Observatory, c/o AURA Inc., Casilla 603, La Serena, Chile\\
$^4$ Universidade do Vale do Para\'iba, Av. Shishima Hifumi, 2911, Cep 12244-000,
S\~ao Jos\'e dos Campos, SP, Brazil\\
$^5$ IATE, Observat\'orio Astron\'omico, Universidad Nacional de 
C\'ordoba, Laprida 854, 5000, C\'ordoba, Argentina}
 
\begin{document}

\date{Accepted -. Received -.}

\maketitle

\label{firstpage}

\begin{abstract}
We present an observational study about the  effects of the interactions in the kinematics, stellar population and
abundances of the components of the galaxy pair AM\,2306-721. Rotation curves for the main and
companion galaxies were obtained, showing a deprojected velocity amplitude of  175 km s$^{-1}$ and
185 km s$^{-1}$, respectively. The interaction between the main and
companion galaxies was modeled using numerical N-body/hydrodynamical simulations, with the result indicating that the current stage of the merger would be about 250 Myr after perigalacticum. 
The spatial variation in the distribution of the stellar
population components in both galaxies was analysed  by fitting
combinations of stellar population models of different age groups. The
central region of main galaxy is  dominated by an old
(5-10\,Gyr) population, while significant contributions from a
young (200\,Myr)  and intermediate (1\,Gyr) components are found in
the disk, being enhanced in the direction of the tidal features. The
stellar population of  the companion galaxy is overall much
younger, being dominated by components with 1\,Gyr or less, quite widely spread over the whole disk. 
Spatial profiles of the oxygen abundance were obtained from the a grid of photoionization
models using the $R_{23}$ line  ratio. The disk of the main galaxy shows a clear radial
gradient, while the companion galaxy presents an oxygen abundance relatively homogeneous across the disk.
The absence of an abundance gradient in the secondary galaxy is
interpreted in terms of mixing by gas flows from the outer parts to the
center of the galaxy due to the gravitational interaction with the
more massive primary.
\end{abstract}


\section{Introduction}
It is widely accepted by some time that merging and interaction
events play an important role in the formation and evolution of
galaxies. Mergers change the mass function
of galaxies, creating a progression from small galaxies to larger
ones; the merging process can also change the morphology of the
constituents, transforming gas-rich spirals in quiescent
ellipticals. Interactions can also trigger a wide  set of physical and
morphological phenomena, such as tidal tails, bridges and shells,
kinematically decoupled cores and star formation enhancements (see review of \citealt{struck99}). 
 
Interacting galaxies also show enhanced star formation when compared
with isolated galaxies. Such enhancement was initially proposed  by \citet{larson78}
to explain the wider range of optical colors found in galaxies in pairs.
Since then, numerous studies have confirmed these results, especially
in the central regions,  through measurements of
optical emission lines  \citep{kennicutt84, kennicutt87, donzelli97, barton03, woods07},  
infrared emission \citep{joseph85, sekiguchi92, geller06} and 
radio  emission \citep{hummel81}. Recent studies have also shown that
this enhancement is a function of the projected galaxy pair separation
(e.g. \citealt{barton00, lambas03, nikolic04}), being
stronger in low-mass  than in high-mass galaxies (e.g. \citealt{woods07,ellison08}). In particular, \citet{ellison08} found that 
the star formation rate as  measured by the H$\alpha$ equivalent width
for galaxies in  pairs selected from the Sloan
Digital Sky Survey, is some  70\% higher when compared to a control sample of galaxies with equal
stellar mass distribution.

There is a connection between the interaction strength and the
morphological distortion in binary galaxies. According to
\citet{mihos96} models, the response of the gas to a close passage 
depends dramatically on the mass distribution of the galaxy, with the
irregularities in the gas velocity field 
tracing the disturbances in the gravitational potential of the
galaxy as observed, for example, in some galaxies in the Virgo
cluster \citep{rubin99}. Combined N-body/hydrodynamic simulations show that 
galaxy-galaxy mergers disturb the gas velocity field significantly, and hence lead to asymmetries and 
distortions in the rotation curves \citep{kronberger06}. However, according
to these authors, no severe distortions are observable about 1 Gyr
after the first encounter. 

The gas motions created by the interaction can also significantly alter the chemical state of the galaxies 
\citep{koeppen90,dalcanton07}, and modify the usually smooth radial
metallicity gradient often found in isolated disk galaxies
\citep{henry99}. Recently, \citet{kewley06} found that O/H abundance
in the central region of nearby galaxy pairs is systematically lower
that that of isolated objects. These authors suggest that the lower metallicity is a 
consequence of gas infall caused by the interaction, but very few
observational studies have been published analysing in detail the
influence of different levels of  interactions in the
metallicity distribution and enrichment properties of galaxies.

Ferreiro \& Pastoriza 2004 (hereafter FP04), in a study of the
integrated photometry and star formation activity in a sample of
interacting systems from the Arp-Madore catalogue \citep{arp87},
found that the galaxies involved have bluer colours than those of isolated galaxies of the same
morphological type, indicating an enhancement of star forming
activity.  This enhancement was also  previously suggested by \citet{donzelli97}  to explain the
slightly larger  values of H$\alpha\, +\, $\ion{N}{ii} equivalent
widths found in these systems, when compared with normal isolated
spiral galaxies. From their sample, we have selected several systems to start
a more comprehensive study of the effects of the interactions in the kinematics, stellar population and
abundances of the galaxies in the so-called ``minor merger'' systems,
defined here as physical pairs with mass ratio 
in the range of $0.04 < M_{secondary}/M_{primary} < 0.2 $.

This paper presents the results for the system AM\,2306-721. This pair is composed by a 
peculiar spiral with disturbed arms  (hereafter, AM\,2306A) in 
interaction with an irregular galaxy (hereafter, AM\,2306B).
Both galaxies contain very luminous  \ion{H}{ii} regions with H$\alpha$ luminosity 
in the range of $8.30\times\,10^{39}\,<\,$L(H$\alpha\,)\, <\,1.27\times\,10^{42}$erg\, s$^{-1}$ 
as estimated from H$\alpha\,$ images \citep{ferreiro08}; 
and high star formation rate in the range of 0.07 to 10  $ M_{\sun}$/yr \citep{ferreiro08}.
The present paper is organized as follows: in Section \ref{datared},  we summarize the observations and data reduction. 
Section \ref{vel} describes the gas kinematics of each galaxy and
Section \ref{numsim} present the numerical
N-body/hydrodynamical simulations of the interaction. Section
\ref{sintese} presents the stellar population synthesis. The
metallicity analysis is  presented in Section \ref{emission}, and the conclusions are summarized in 
Section \ref{final}.

\section{Observations and data reduction}
\label{datared}

Long slit spectroscopic data was obtained in 20/21
June 2006 and 23/24 June 2007 with the Gemini Multi-Object
Spectrograph at Gemini South, as part of poor weather programmes
GS-2006A-DD-6 and GS-2007A-Q-76. Spectra in the range 3\,350 to
7\,130\AA\ were taken in two settings with the B600 grating, and the 1$\arcsec$ slit, yielding a 
spectral resolution of 5.5\AA. The frames were binned on-chip by 4 and 2 pixels in
the spatial and spectra directions, respectively, resulting in a
spatial scale of 0.288 $\arcsec$\,pxl$^{-1}$, and 0.9\AA\,pxl$^{-1}$
dispersion.

\begin{figure}
\centering
\includegraphics*[width=\columnwidth]{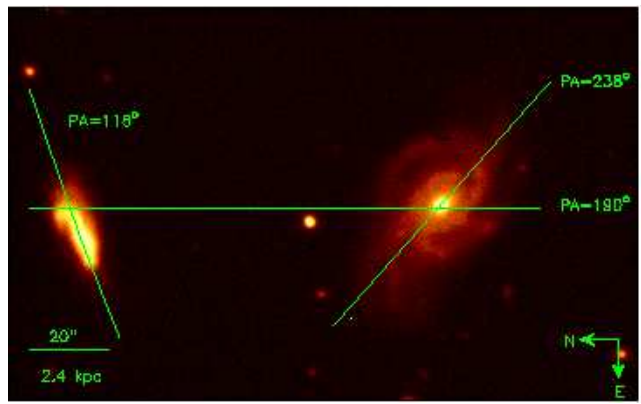}  				     
\caption{GMOS-S $r'$-band image of AM\,2306-721 with the observed slit
positions.}
\label{field}
\end{figure}

Spectra were taken at three different position angles on the sky:
PA=238$\degr$ and PA=118$\degr$ corresponding to positions along the
major axis of AM\,2306A and AM\,2306B, respectively; and
PA=190$\degr$, a position cutting across the
disk of both galaxies. The exposure time on each single frame was limited to 700 seconds to 
minimize the effects of cosmic rays, with four frames being obtained
for each slit position to achieve suitable signal. The slit positions are shown in Fig. \ref{field}, 
superimposed on the $r'$-band image of the pair. Table \ref{observ}
gives the journal of observations. Conditions during both runs were
not photometric, with thin cirrus and image quality in the range 1.0$\arcsec$
to 2.5$\arcsec$ (as measured from stars in the acquisition images taken
just prior to the spectroscopic observations).

\begin{figure*}
\includegraphics*[angle=-90,width=\textwidth]{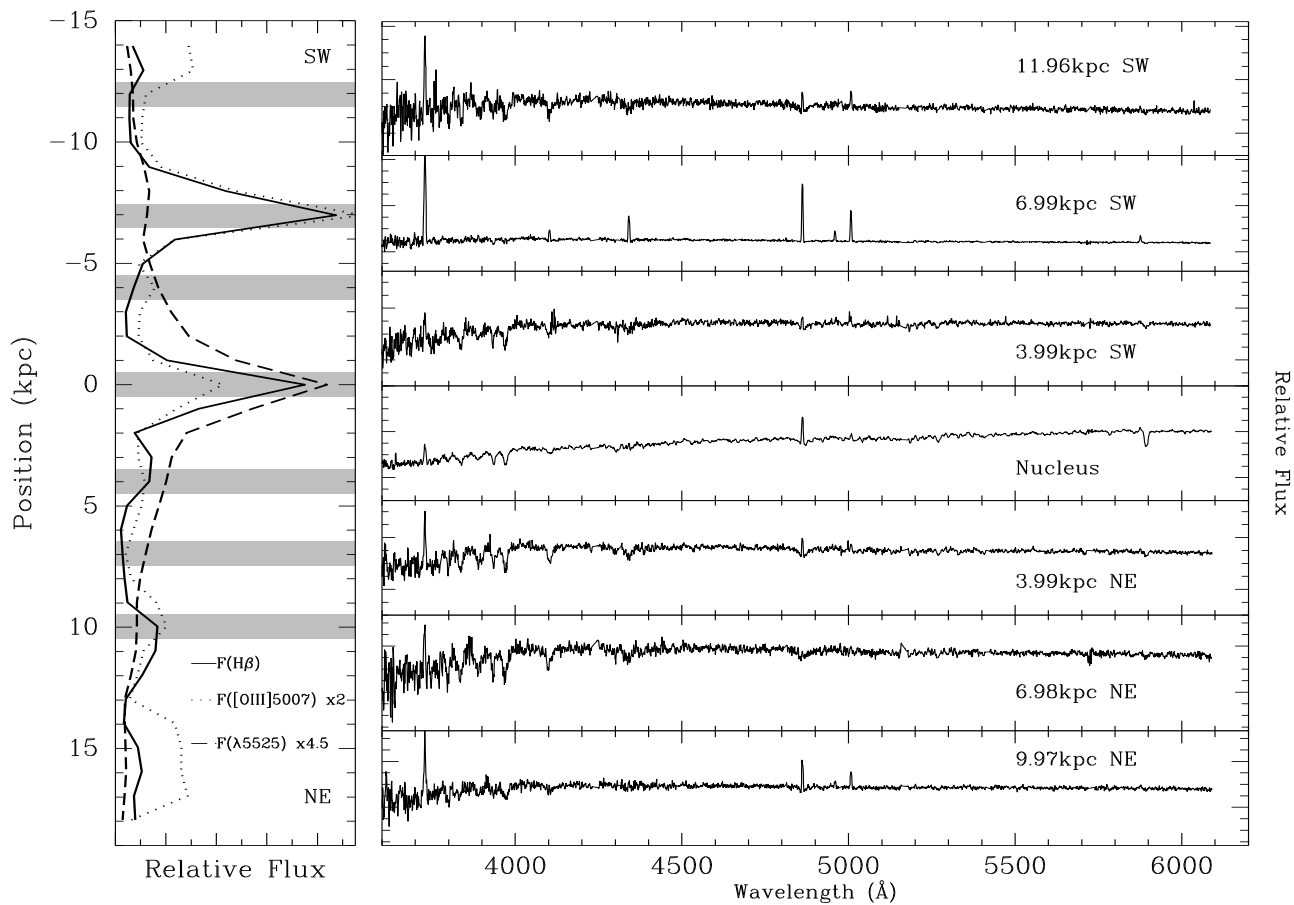}				     
\caption{Spatial profiles of H$\beta$,
[\ion{O}{iii}] $\lambda 5007$ and  $\lambda 5525$ flux 
along PA=238\degr\, for AM\,2306A (left); and a sample spectra in the range of 3600 to 6000 \AA\,
from different regions along PA=238\degr\, for AM\,2306A 
as marked in the shaded areas in the spatial profiles  (right). The spectra have been normalized
at $\lambda\,5870$ \AA\, and are plotted in the rest frame wavelength.
}
\label{spectraa}
\end{figure*}

\begin{figure*}
\includegraphics*[angle=-90,width=\textwidth]{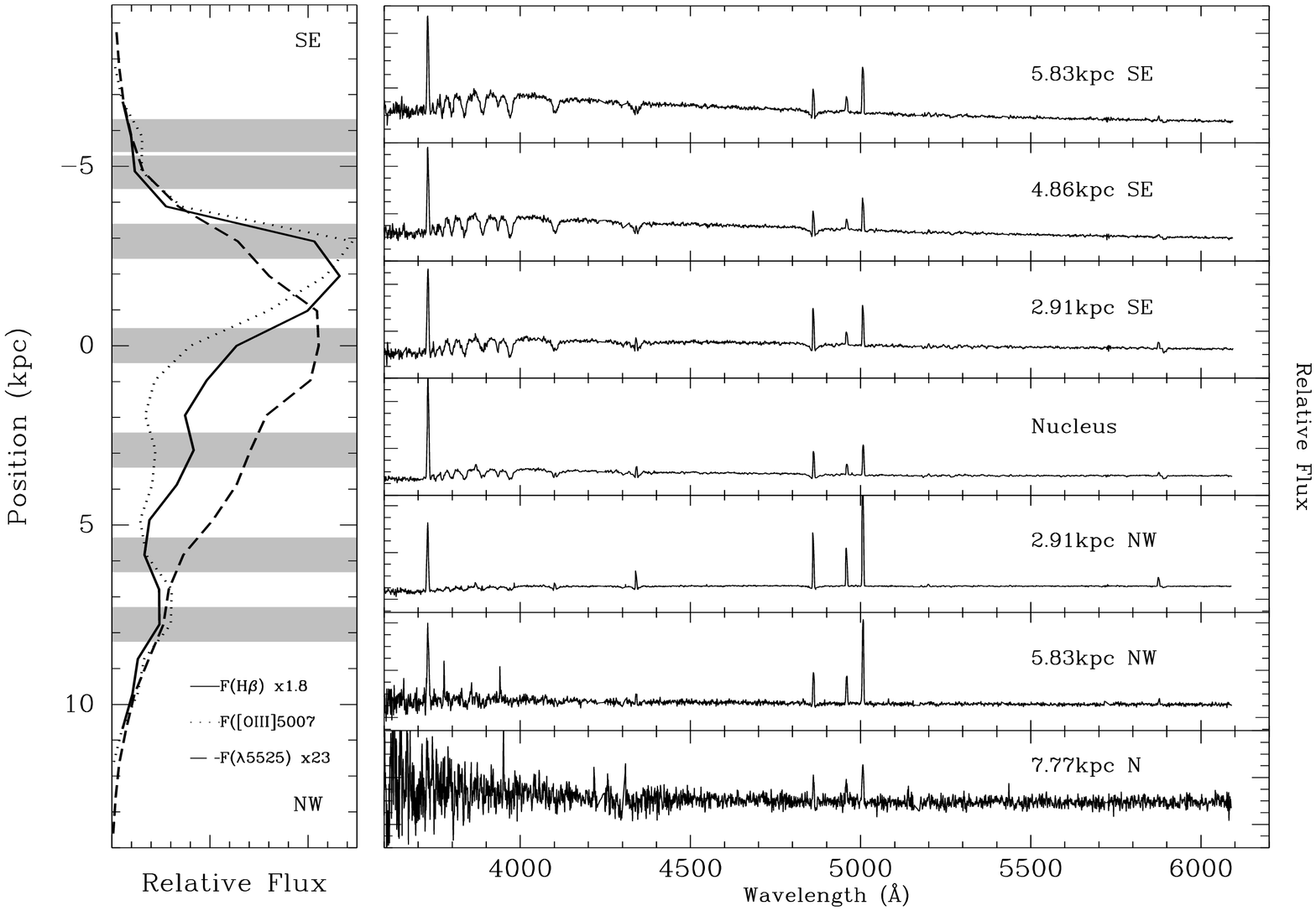}					     
\caption{Same as Fig.~\ref{spectraa}, but along PA=118\degr\, for AM\,2306B}
\label{spectrab}
\end{figure*}

The spectroscopic data reduction was carried out  using 
the {\sc gemini.gmos} package as well as  generic {\sc IRAF} tasks. We followed the standard procedure for
bias correction, flat-fielding, cosmic ray cleaning, sky subtraction,
wavelength and relative flux calibrations. In order to increase the
signal-to-noise ratio, the spectra were then extracted summing over
six rows. Each spectrum thus represents an aperture of 1$\arcsec \times
1.73 \arcsec$. For a distance of  119\,Mpc for  AM\,2306A, and
116\,Mpc for AM\,2306B, (estimated from the radial velocities derived in Section  \ref{vel} and   
adopting $H_0$=75\,km\,s$^{-1}$\,Mpc$^{-1}$), this aperture
corresponds to a region of 577\,$\times$\,998\,pc$^2$ for  AM\,2306A; and
562\,$\times$\,972\,pc$^2$ for  AM\,2306B. The nominal centre of each
slit was chosen to be the continuum peak at $\lambda\,5525\AA$. Fig.~\ref{spectraa} and ~\ref{spectrab}  shows a sample 
of spectra of AM\,2306A  and AM\,2306B extracted along the slit for
PA=238$\degr$ and PA=118$\degr$, respectively.

\begin{table}
\caption{Journal of observations}
\label{observ}
\begin{tabular}{lccc}
\noalign{\smallskip}
\hline
\noalign{\smallskip}
Date (UT) & Exposure time(s) & PA (\degr)& $\Delta \lambda (\AA)$ \\
\noalign{\smallskip}
\hline
\noalign{\smallskip}
2006/06/20 &  4$\times$700    & 190       & 4280-7130  \\
2006/06/20 &  4$\times$700    & 238       & 4280-7130  \\
2007/06/24 &  4$\times$700    & 190       & 3350-6090  \\
2007/06/24 &  4$\times$700    & 238       & 3350-6090  \\
2007/06/24  &  4$\times$700    & 118       & 3350-6090  \\
\noalign{\smallskip}
\hline
\noalign{\smallskip}
\end{tabular}
\end{table}

\begin{figure*}
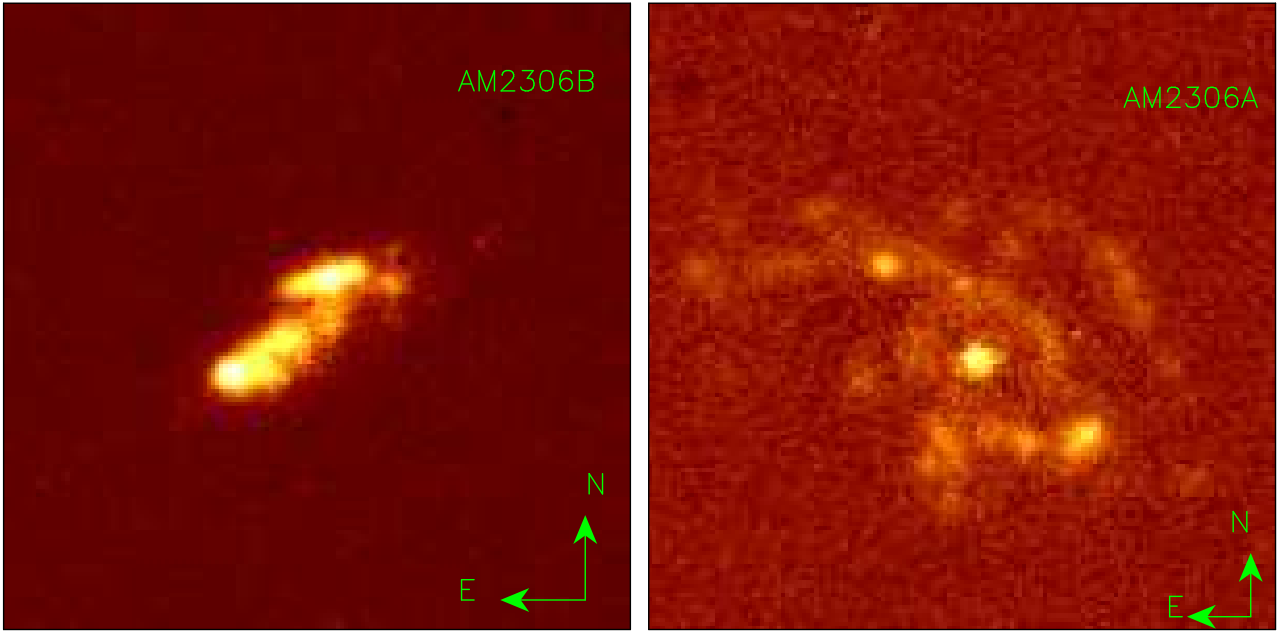

\subfigure{
\includegraphics*[angle=-90,width=\columnwidth]{am2306b.eps}}					     
\subfigure{
\includegraphics*[angle=-90,width=\columnwidth]{am2306a.eps}}					     
\caption{H$\alpha$ image of AM\,2306B (left) and AM\,2306A
(right). Description and detailed analysis of these images have been
presented in FP04.}
\label{haimage}
\end{figure*}

\section{Ionized gas kinematics}
\label{vel}
The radial velocity was estimated from the strongest emission lines present in the spectra, namely 
$\rm H\gamma$ $\lambda 4340$, $\rm H\beta$ $\lambda 4861$,[\ion{O}{iii}] $\lambda 5007$, 
$\rm H\alpha$ $\lambda 6563$, [\ion{N}{ii}] $\lambda 6584$, and [\ion{S}{ii}] $\lambda 6717$.
The final radial velocity for each spectrum was obtained by averaging
the individual measurements from the detected emission lines, and the errors estimated from the
standard deviation of these measurements around that mean.

Using the acquisition images in the $r'$ filter, we did a simple
isophote fitting using the {\sc stsdas.ellipse} task to obtain the
position angle of the major and minor axes of both galaxies. We found
that the major axis for AM2306A is at PA=$236\degr$, and for AM2306B at
PA=$108\degr$. The inclination of each galaxy  with respect to the
plane of the sky was also computed as $\cos(i)=b/a$, where $a$ and $b$ are
the minor and major semi-axes of the galaxy, respectively. We obtained
$i=56\degr$ for the main galaxy and  $i=60\degr$ for the secondary, in
agreement with the values measured by FP04.

\begin{figure*}
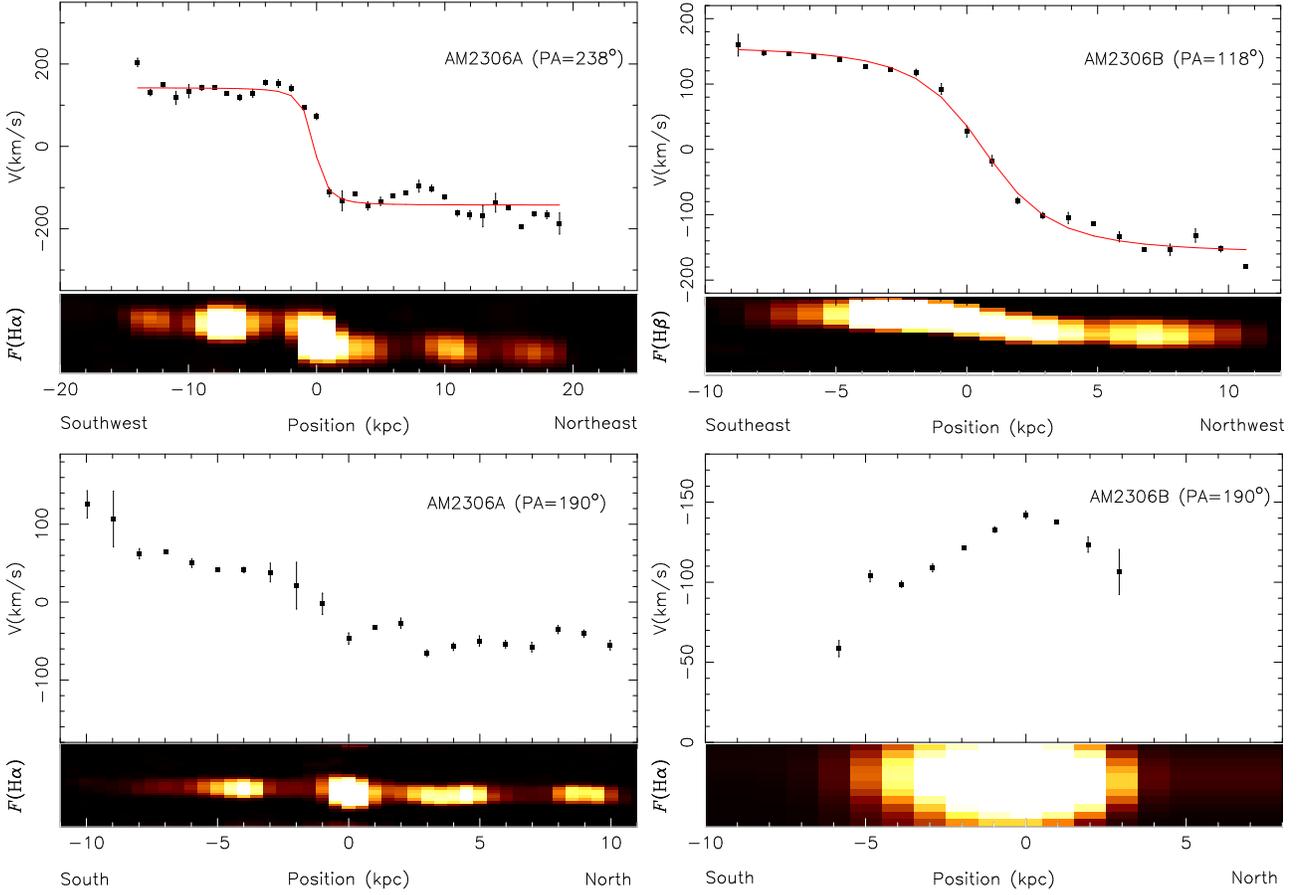

\subfigure{
\includegraphics*[angle=-90,width=\columnwidth]{vel_a.eps}}					     
\subfigure{
\includegraphics*[angle=-90,width=\columnwidth]{vel_bf.eps}}					     
\subfigure{
\includegraphics*[angle=-90,width=\columnwidth]{vel_a1.eps}}					     
\subfigure{
\includegraphics*[angle=-90,width=\columnwidth]{vel_b1.eps}}					     
\caption{Observed mean radial velocity and 2D spectra of H$\alpha$ and H$\beta$  
emission lines as a function of apparent galactocentric distance. The
left panels show the kinematics along PA=238\degr (top) and
PA=190\degr (bottom) of AM\,2306A, respectively. The right panels
show the kinematics along PA=118\degr (top), and PA=190\degr (bottom)
of AM\,2306B, respectively. The velocity scale correspond to the
observed and model values after subtraction of the systemic velocity
of each galaxy, without correction by the inclination in the plane of
the sky.}
\label{radial_velocity}
\end{figure*}

The rotation curves along the observed slit positions for both galaxies
are presented in Figure \ref{radial_velocity}, as well as the
coresponding two-dimentsional spectra of H$\alpha$ and/or H$\beta$
emission lines. The radial velocity values after subtraction of the
systemic velocities (as determined from the rotation model described
below) are listed in Table \ref{velo}.

Both galaxies show a fairly symmetric rotation curve, so we  adopted a very simple approximation 
for the observed velocity distribution, assuming that the gas moves under a
logarithmic gravitational potential, following circular 
orbits close to a plane $P(i,\psi_{0})$, characterized by its inclination 
to the plane of the sky $(i)$ and the position angle (PA) 
of the line of nodes  $\psi_0$. This assumption results in an observed radial circular velocity
$v(r,\psi)$ in the plane of the sky given by \citet{bertola91}:

\begin{equation}
v(r,\psi)= V_{s} + \frac{V_{0}R \cos(\psi-\psi_{0})\sin(i)\cos(i)}{\sqrt{R^{2}\eta + R_{c}^{2}\cos^{2}(i)}},
\label{v_mol}
\end{equation}
with
\begin{equation}
\eta \equiv [\sin^{2}(\psi-\psi_{0}) + \cos^{2}(i)\cos^{2}(\psi-\psi_{0})],
\label{v_molcont}
\end{equation}
where $V_{s}$ is the systemic velocity, $R$ is the radius in the plane of 
the galaxy and $V_{0}$ and $R_{c}$  are parameters that define 
the amplitude and shape of the curve. The fit of the rotation curves for each galaxy are shown in Fig. \ref{radial_velocity} and the parameters
obtained are listed in Table \ref{par}.

\begin{table}
\caption{Kinematical parameters}
\label{par}
\begin{tabular}{lccc}
\noalign{\smallskip}
\hline
\noalign{\smallskip}
& \multicolumn{1}{c}{AM\,2306A (PA=238\degr)} & &\multicolumn{1}{c}{AM\,2306B (PA=118\degr)}\\
\noalign{\smallskip}
\cline{2-2}
\cline{4-4}
\noalign{\smallskip}
Parameter & Model &   & Model \\
\noalign{\smallskip}
\hline
\noalign{\smallskip}
$V_{s}$ (km/s)  & 8919$\pm$5     & &8669$\pm$3 \\
$V_{0}$ (km/s)  & 175$\pm6$          & &185 $\pm$6 \\
$R_{c}$ (kpc)   & 1$\pm$0.3            & &1.6 $\pm$0.2 \\
\noalign{\smallskip}
\hline
\noalign{\smallskip}
\end{tabular}
\end{table}

\begin{table*}
\caption{Radial velocities. The full table is available as Supplementary Material to the online version 
of this article from http://www.blackwell-synergy.com.}
\centering
\label{velo}
\begin{tabular}{rrrrrrrrrrr}
\noalign{\smallskip}
\hline
\hline
\noalign{\smallskip}
 \multicolumn{5}{c}{AM\,2306A} &  &\multicolumn{5}{c}{AM\,2306B}\\
\cline{1-5}
\cline{7-11}
\noalign{\smallskip}
\multicolumn{2}{c}{PA=238$^{o}$} & &\multicolumn{2}{c}{PA=190$^{o}$} &   
  & \multicolumn{2}{c}{PA=118$^{o}$}& & \multicolumn{2}{c}{PA=190$^{o}$}\\ 
\noalign{\smallskip}
\cline{1-2}
\cline{4-5}
\cline{7-8}
\cline{10-11}
\noalign{\smallskip}
\multicolumn{1}{c}{R (kpc)}  &\multicolumn{1}{c}{V (km/s)} & & \multicolumn{1}{c}{R (kpc)} &\multicolumn{1}{c}{V (km/s)} & &\multicolumn{1}{c}{R (kpc)} &\multicolumn{1}{c}{V (km/s)}& &\multicolumn{1}{c}{R (kpc)} &\multicolumn{1}{c}{V (km/s)}\\
\noalign{\smallskip}
\hline
\noalign{\smallskip}
13.96 SW     &  204$\pm$10   & & 13.96 S     & ...~~~~    &  &...~~~~    &  ...~~~~         &&...~~~~    & ...~~~~  \\
12.96 SW     &  131$\pm$8    & & 12.96 S     & ...~~~~    &  &...~~~~    &  ...~~~~	    &&...~~~~    & ...~~~~   \\
11.96 SW     &  150$\pm$5    & & 11.96 S     & ...~~~~    &  &...~~~~    &  ...~~~~	    &&...~~~~    & ...~~~~   \\
10.97 SW     &  118$\pm$16   & & 10.97 S     & ...~~~~    &  &...~~~~    &  ...~~~~         &&...~~~~    & ...~~~~    \\
9.97  SW     &  134$\pm$16   & & 9.97  S     &125$\pm$18  &  &...~~~~    &  ...~~~~         &&...~~~~    & ...~~~~     \\
\noalign{\smallskip}
\hline
\hline
\noalign{\smallskip}
\end{tabular}  
\end{table*}

\begin{table}
\caption{Parameters used on the simulations}  
\label{tab:modelpars}
\begin{small}
\begin{tabular}{l c c}
\hline
                                &{\bf AM\,2306A}	&{\bf AM\,2306B}	\\
\hline
Number of points in disk        & 2048     		&   2048		\\
Disk mass               	& 0.5      		&   0.17		\\
Disk radial scale length        & 0.6      		&   0.6 		\\
Disk vertical scale thickness   & 0.06       		&   0.05		\\
Reference radius R$_{ref}$      & 0.06        		&   1.2			\\
Toomre Q at R$_{ref}$           & 1.5       		&   1.5			\\
\hline
Number of points in gas disk    & 2048			&   2048		\\
Gas disk mass               	& 0.05       		&   0.017		\\
Gas disk radial scale length    & 1.0			&   0.75		\\
Gas disk vertical scale thickness	& 0.02      	&   0.03		\\
Toomre Q at R$_{ref}$           & 1.5       		&   1.5			\\
\hline
Number of points in bulge       & 256       		&   256 		\\
Bulge mass              	& 0.08			&   0.01		\\
Bulge radial scale length       & 0.1       		&   0.06		\\
\hline
Number of points in spherical halo  & 2048	     	&   2048		\\
Halo mass               	& 3.5       		&   1.7			\\
Halo cutoff radius          	& 8.0       		&   4.0			\\
Halo core radius            	& 2.5        		&   0.6			\\
\hline
\smallskip
\end{tabular}
{\bf Notes:} Simulations were done in a system of units with
G=1. Model units scale to physical ones as: unit length is
3.5\,kpc, unit velocity is 262\,km\,s$^{-1}$, unit mass is
$5.586\times\,10^{10}\,\mathrm{M}_\odot$ and unit time is 13.062\,Myr.
\end{small}
\end{table}

The heliocentric  velocity of the main galaxy is found to be 8\,919 km s$^{-1}$.
The rotation curve obtained for AM\,2306A along PA=238\degr\ is well
represented by the simple model above. This rotation curve is typical
of spiral disks, rising shallowly and flattening at an observed amplitude of
145 km\,s$^{-1}$. As can be seen in Fig.\ref{radial_velocity}, there
are pronounced velocity features indicating local
deviations from a smooth rotational field. One of such well defined
features is located towards the northeast along PA=238\degr, at about
8 kpc projected distance from the nucleus, with an amplitude
(peak-to-peak) of about 100 km\,s$^{-1}$. This feature is located
where the slit crosses the the outer spiral arm  as seen in the $\rm
H\alpha$ image (Fig. ~\ref{haimage}, right). This image also makes more
obvious the disturbed spiral structure of the main component, very
likely due to the interaction with AM\,2306B. 

For the secondary galaxy, the rotation curve model results in an
heliocentric velocity of 8\,669 km\,s$^{-1}$. The observed radial
velocities  along PA=118\degr\ are also very well represented 
by the model. The rotation curve is quite similar  to that of AM\,2306A, rising shallowly 
and flattening at an observed amplitude of about 158 km
s$^{-1}$. Although the young star forming population of the secondary
galaxy show clear morphological irregularities (Fig. ~\ref{haimage},
left), the rotation curve does not present significant deviations from
the smooth representation of the velocity field.

\citet{donzelli97} estimated systemic velocities of 8\,762 and 9\,069
km\,s$^{-1}$ for AM2306A and AM2306B, respectively.  Following our
results, we verified that the values presented by those authors were
reversed, and therefore our estimates are found to agree to within 
2 \% with those previous determinations.

An estimate of the dynamical mass can be derived by assuming that 
the mass inside a certain radius is given by $ M(R)=RV^{2}/G$. 
For the main galaxy, the deprojected velocity amplitude is 175 km s$^{-1}$ 
and its dynamical mass is $ 1.29 \times 10^{11} M_{\sun}$
within  a radius of 18 kpc. For the companion galaxy, the
deprojected velocity amplitude is 185 km s$^{-1}$ and the
dynamical mass within a radius of  10.7 kpc is $ M(R)= 8.56 \times 10^{10} M_{\sun}$. It is important to emphasize that
the maximum radius to which we can observe the gas in emission is
quite certainly smaller than the total radius of the galaxies, so
our estimates of the dynamical mass give a lower limit to the
actual dynamical mass of each system.

In determining the dynamical mass,  the main source of error is in the
estimation of the deprojected velocity, which in turn is highly
dependent on the assumed inclination of the galaxies with respect to
the plane of the sky. For AM\,2306A the inclination angle is well
determined from the isophote fitting, with an estimated error of about
1\%, propagating to a 2\% uncertainty in the resulting total
mass. For AM\,2306B, the irregular morphology of the disk in optical
images implies that the calculated value of the inclination angle
varies significantly with radius. In this case, if we considered the isophote fitting from the 
regions more perturbed of AM\,2306B to estimate the inclination angle, a maximum angle inclination of about 
12 \% would be obtained, implying in a reduction of up to 18 \%  in the total mass.

\section{Numerical Simulations}
\label{numsim}
In order to reconstruct the history of the AM\,2306-721 system and to
predict the evolution of the encounter, we modeled the interaction
between AM\,2306A and AM\,2306B through numerical simulations using
the the N-body/SPH code GADGET-2 developed by \citet{springel05}. The galaxies were modeled following the
prescription of \citet{hernquist93}, including  a gaseous disk
component. The observed morphology and rotation curves presented in
Sec. ~\ref{vel} provide the constraints to the simulations. 

As for any study of this kind, recreating the evolution of the
AM\,2306-721 system requires solving the reverse problem of calculating
the orbit followed by the two galaxies from their observed
properties. This, like most similar cases, is not a fully determined problem, since the observations do not provide all
the necessary information to uniquely identify the
solution. Therefore, in order to restrict the parameter space when
setting up the initial conditions for the simulations, we first calculate orbits that 
satisfy the constraint given by the observed radial velocity
difference, testing different eccentricities, pericenter  and
line-of-sight direction distances. From that subset, based on the
observed morphology, we select a few orbits to run the full
simulations, from which the one that best fits the observed properties
is selected. 

\begin{figure*}
\centering
\includegraphics*[width=\columnwidth]{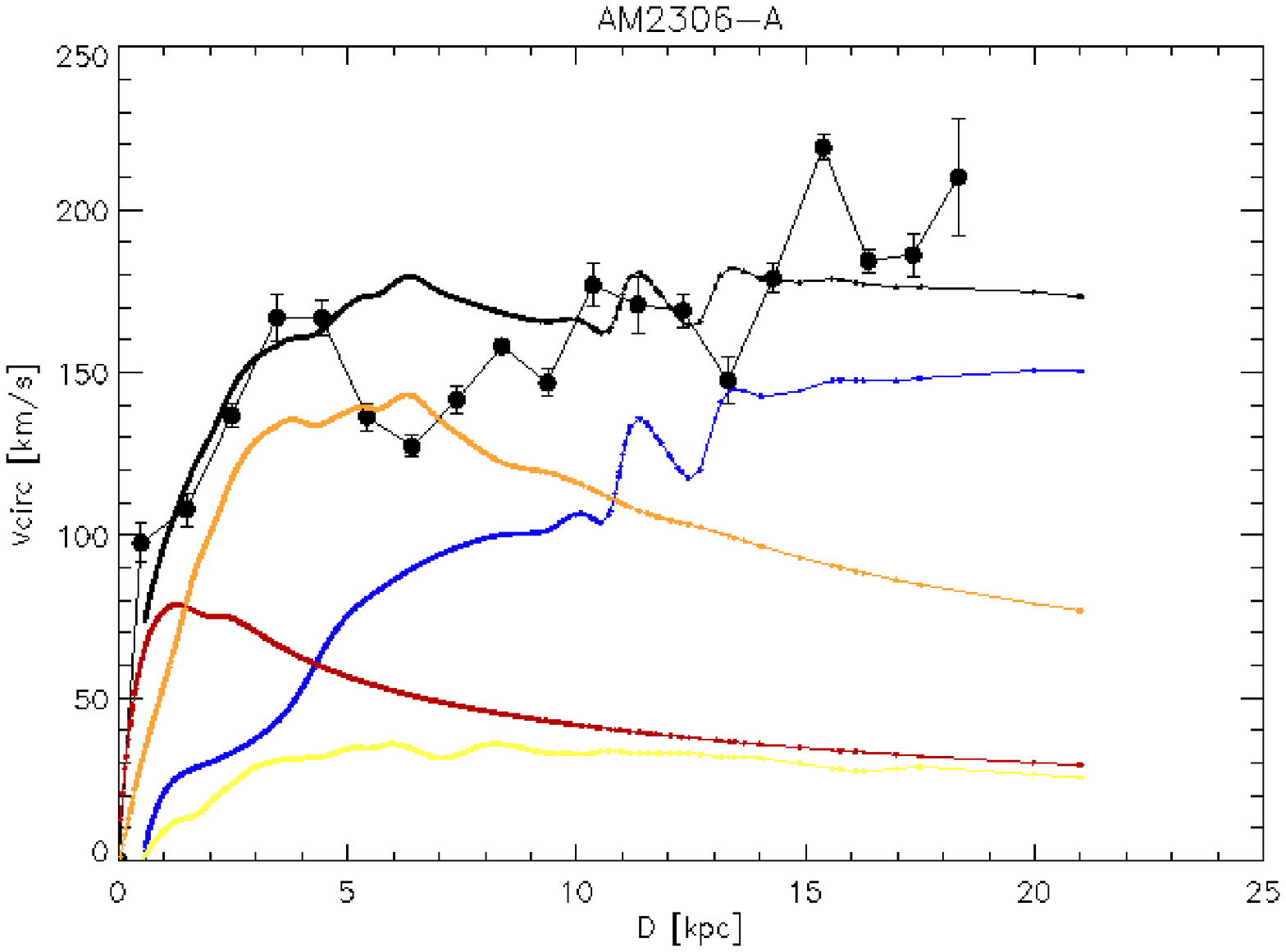}
\includegraphics*[width=\columnwidth]{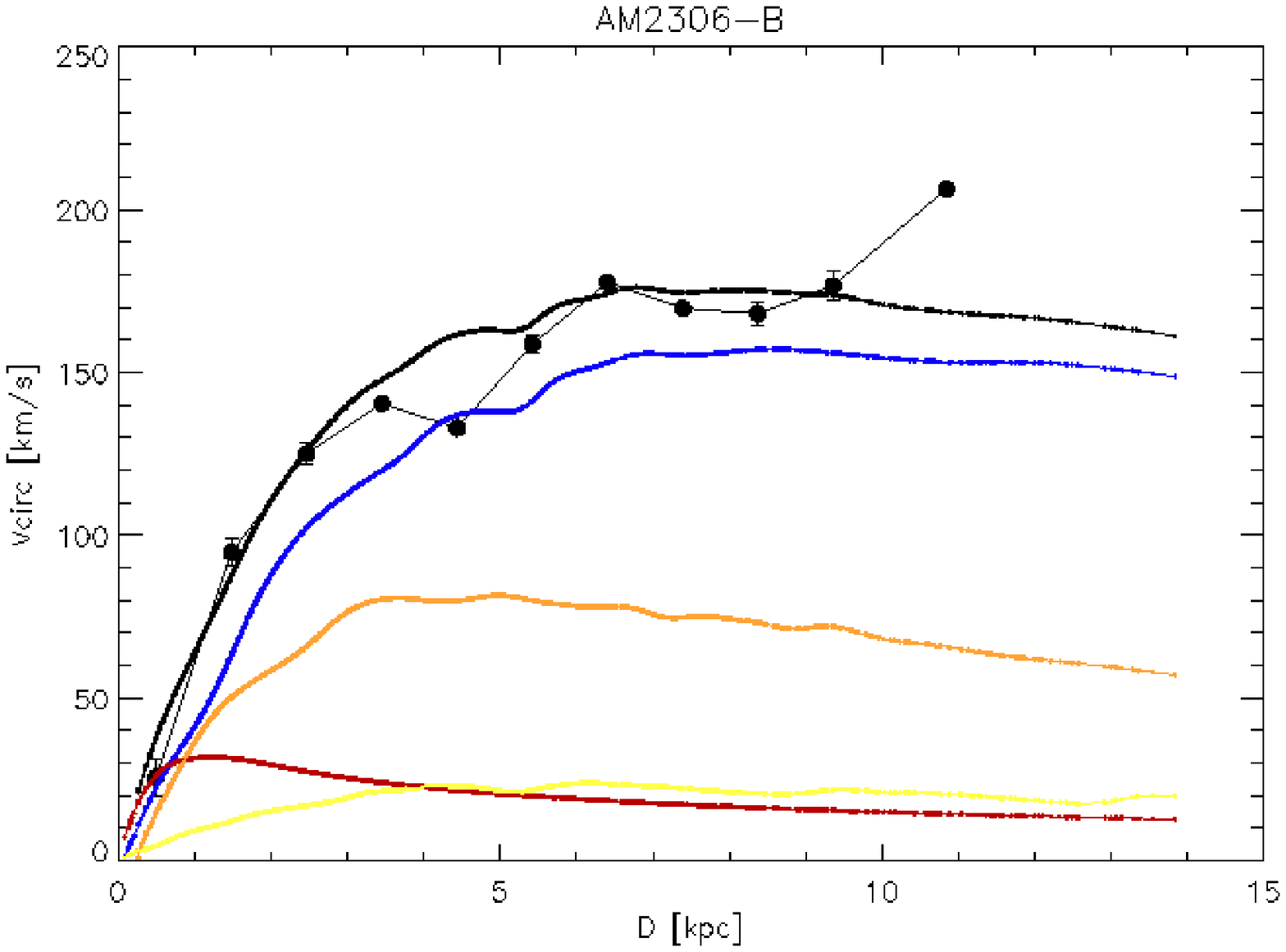}
\caption{Circular velocity curves for galaxy models AM\,2306A (left)
and AM\,2306B (right). The rotation curves of the individual model
components are shown: halo in orange, bulge in red, stellar disk in
blue and gas disk in yellow. The composite rotation curve is the
continuous thick black line. Points with error bars connected by thin
black lines are the observed radial velocity curves, folded around the
rotational centre and  corrected by the inclination of each galaxy.}
\label{vcirc}
\end{figure*}

The orbit that best reproduces the observational properties is found
to be hyperbolic, with an eccentricity $e=1.15$ and perigalacticum of
$q=5.25$ kpc .
The orbital plane is perpendicular to the plane of the sky, and
intersects the later in the line that connects AM\,2306A and AM\,2306B
as projected in the sky. In perigalacticum, AM\,2306B was behind
AM\,2306A, along the line of sight. The parameters for the best fit
model are presented in Table~\ref{tab:modelpars}. The composite
rotation curve, as well as the individual model components and observed circular
velocity, corrected by the inclination of each galaxy as given in
Section~\ref{vel}, are shown in Figure~\ref{vcirc}. The models indicate a total
mass of $1.82\times10^{11} M_{\sun}$ for AM\,2306A and of
$8.68\times10^{10} M_{\sun}$ for AM\,2306B. The mass of each individual component
(disk, bulge, halo and gas) is  quoted in
Table~\ref{tab:modelpars}.

\begin{figure*}
\includegraphics*[width=3.87cm]{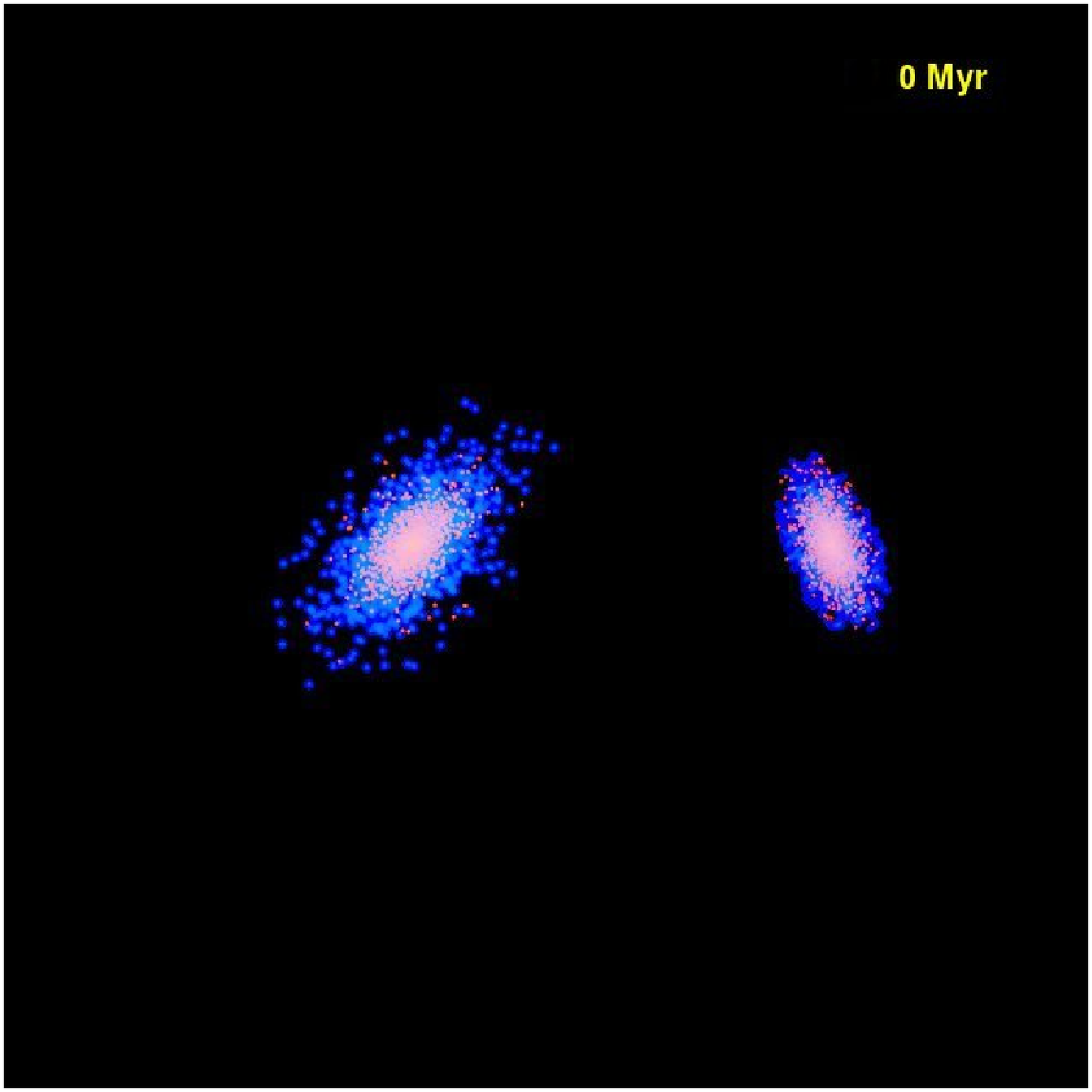}
\includegraphics*[width=3.87cm]{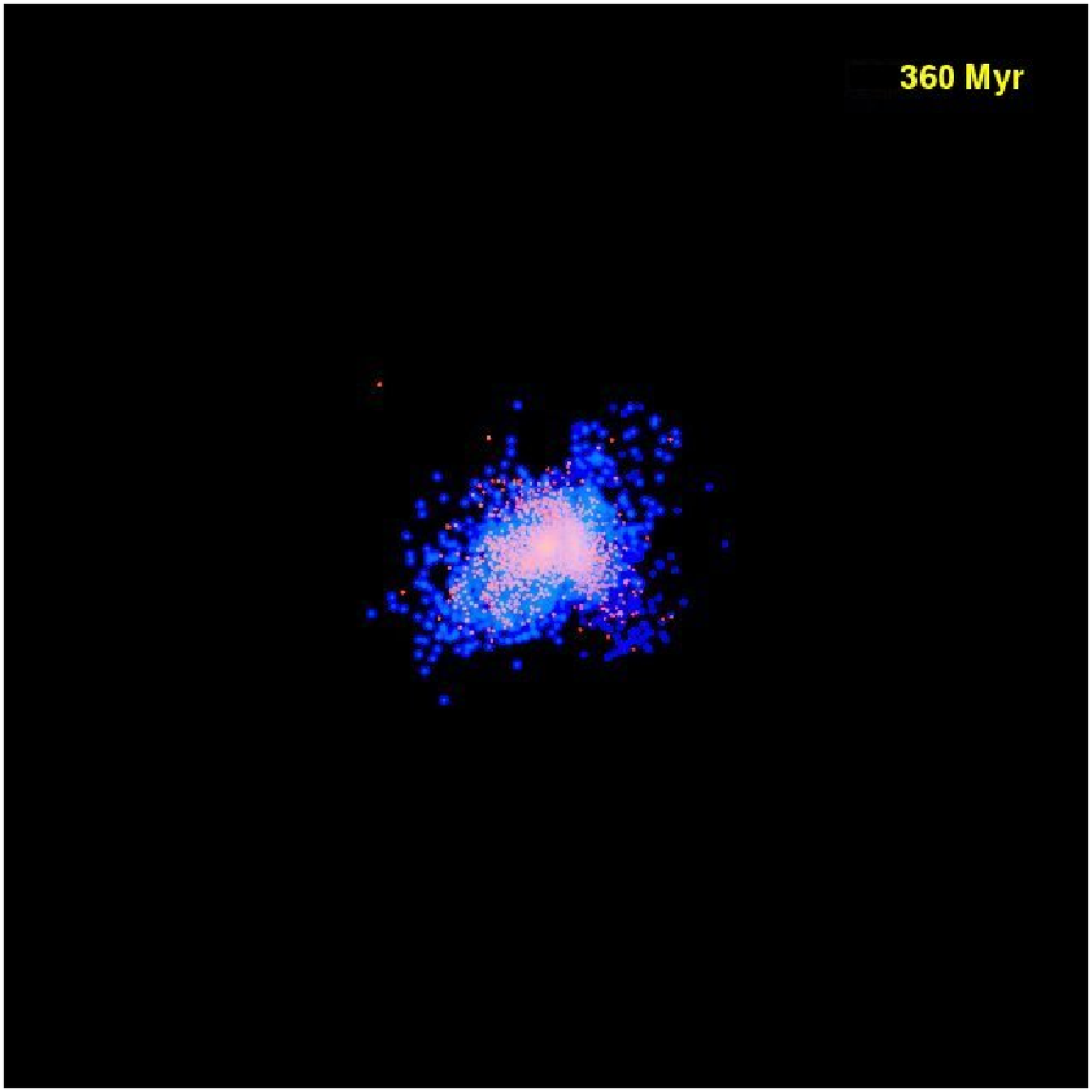}
\includegraphics*[width=3.87cm]{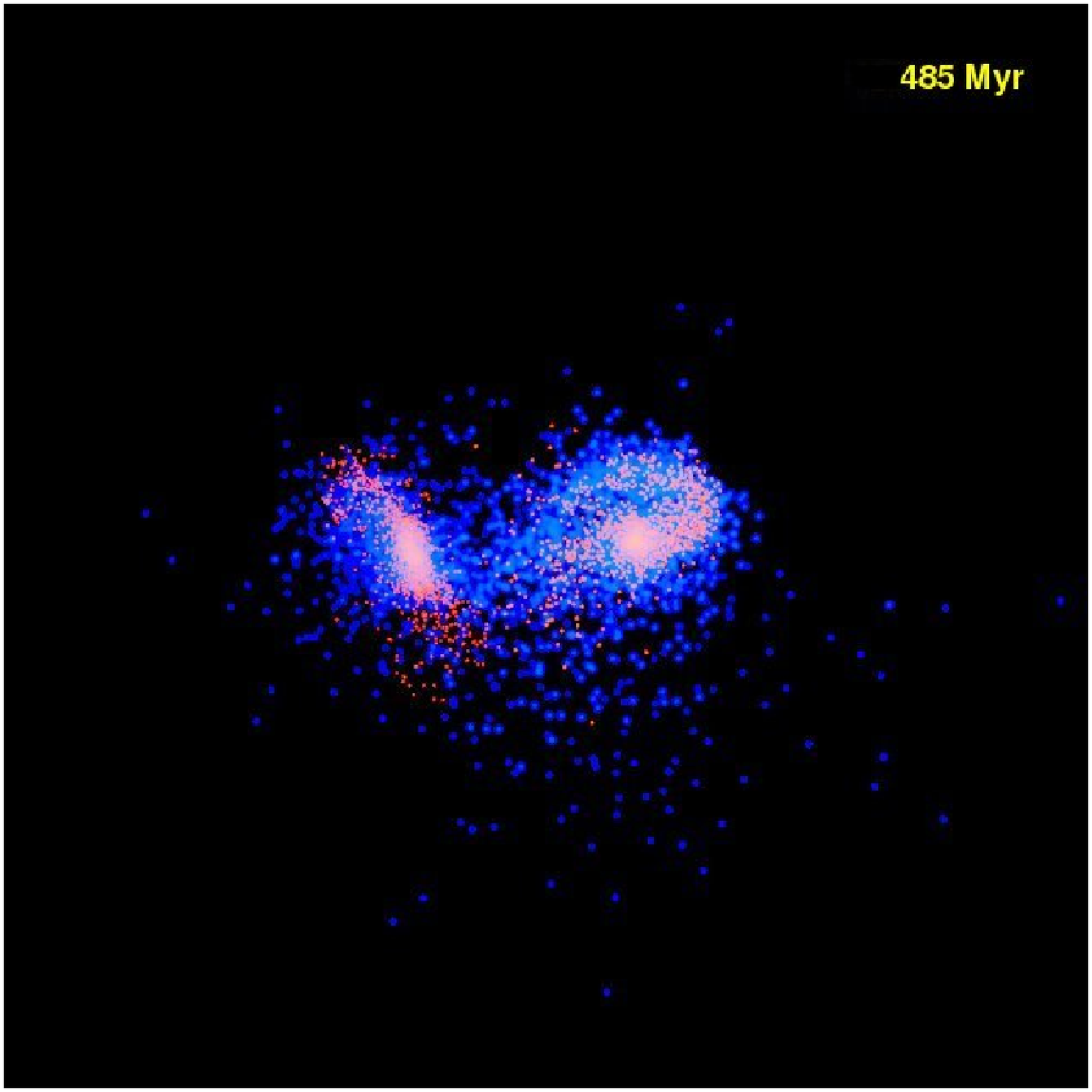}
\includegraphics*[width=3.87cm]{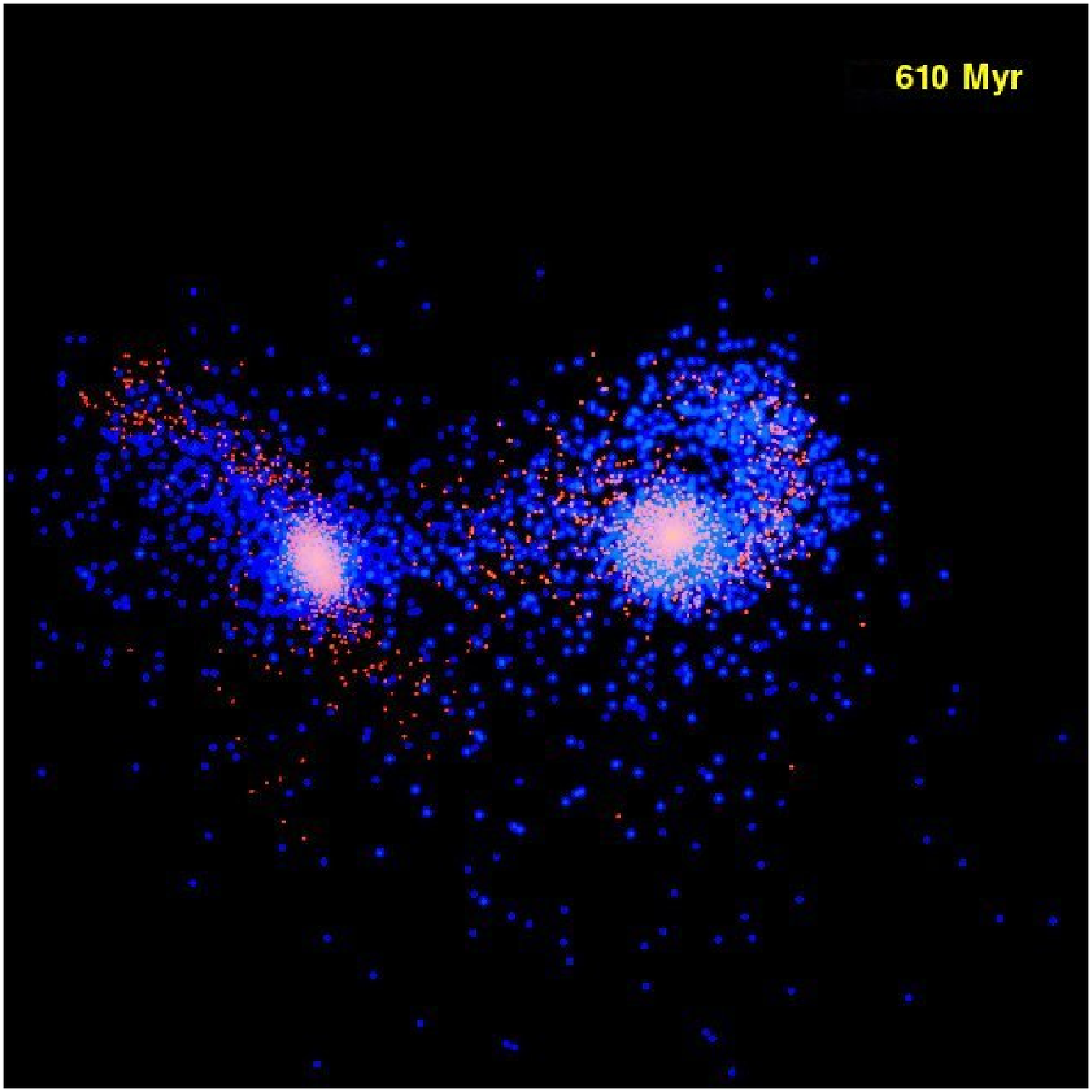}
\caption{Four snapshots representing the evolution of the
encounter. Time is shown in Myr in the upper right corner of each
frame, with respect to the beginning of the simulation. Stars are
plotted in red, gas in blue. Orbital perigalacticum occurs at $t=360$
Myr (upper right frame). Best fit to current evolutionary stage of the
AM\,2306-721 system is about 250 Myr after perigalacticum (bottom
right frame).}
\label{simul}
\end{figure*}

Figure~\ref{simul} shows the time evolution of the
encounter. Simulation starts 360 Myr before perigalacticum, and the time is shown in
Myr in the upper right corner of each frame.  The situation
that best reproduces the morphology and kinematics of AM\,2306-721
system at the current stage is  $t=610$ Myr, or about 250 Myr after
perigalacticum. The overall large scale morphology and kinematics
agree well with observations, considering the low resolution of the
simulation.

\section{Stellar Population Synthesis}
\label{sintese}
A detailed study of the star formation in minor merger galaxies is an
important source of information not only on the age distribution of
their stellar population components, but to better understand several
aspects related to the interacting process, its effect in the
properties of the individual galaxies and their later evolution. The
absorption features arising from the  stellar component also affect to
different degrees the measured intensity of the emission line in the
spectrum of the gaseous component. This effect is more prominent in,
but not restricted to, the Balmer lines, so the stellar population
contribution must be subtracted from each  spectra in order to
investigate the physical properties of the gas in these galaxies.

To investigate the star formation history of AM\,2306A and AM\,2306B
we use the stellar population synthesis method developed by \citet{bica88}. This method employs 
the equivalent widths
$W_{\lambda}$ of several spectral absorption features and the
measured continuum fluxes $F_{\lambda}$  at different wavelengths,
comparing  them to those of a model computed from a base of simple
stellar population (SSP) elements with known ages and metallicities. 
The algorithm  generates all possible combinations 
of the base elements according to a given flux
contribution step and compares the resulting $W_{\lambda}$ and
continuum points to the input ones.  The allowed solutions are those
which reproduce, within predefined limits, the observed $W_{\lambda}$
and $F_{\lambda}$. All these possible solutions are then averaged, and
this average is adopted as the final synthesis, with  the uncertainty
for each individual age component given by the standard deviation of
the contribution from each allowed  solution around the mean adopted
value. The code used here is based on an upgraded version  of the one
presented by \citet{schmitt96}, and includes a correction for the effect of the internal extinction.

The spectral windows for measuring  $W_{\lambda}$ are those defined in
the Lick system \citep{worthey97,trager98}, with the addition of the K
\ion{Ca}{ii} (corresponding to the spectral window
$\lambda\,\lambda\,3925-3945\AA$)),   and H11 $\lambda\,3770\AA$ lines  (spectral window
$\lambda\,\lambda\,3765-3785\AA$). Previous to the measurements of  the $W_{\lambda}$ and $F_{\lambda}$ the spectra were 
corrected by the foreground (Galactic) reddening of $E(B-V)=0.03$ mag taken from \cite{schlegel98};  
and  normalized to $\lambda\,5870$\AA. The $W_{\lambda}$ were measured using the 
{\it PACCE} code kindly provided by \cite{vale07}. The  $W_{\lambda}$ values measured for each 
galaxy at different galactocentric distances and position angles are listed 
in Table \ref{eqw}. 

The SSP base was created from the \citet{bruzual03} evolutionary 
stellar population models, which are based on a high resolution library of observed stellar 
spectra. This library allows us to derive detailed spectral evolution of  simple 
stellar populations across the wavelength range of  3\,200 to 9\,500 $\AA$ with a wide range of 
metallicities. We used the Padova 1994 tracks as recommended by \citet{bruzual03}, with the initial mass function 
of Salpeter \citep{salpeter55}. The final base contains five spectra
corresponding to SSPs with ages of 2.5 Myr, 200 Myr, 1 Gyr, 5 Gyr and
10 Gyr, and solar metallicity. 
The adopted number of elements in the
base was defined by a compromise between having enough age resolution,
having a limited number of features available to constrain the fit, and the desire to mantain a consistent base for all
apertures in both galaxies. For the few positions with higher S/N
spectra in each galaxy, we tested a more detailed age grid, including components with ages 2.5 Myr, 100 Myr, 200 Myr, 500 Myr
1 Gyr, 5 Gyr and 10 Gyr, and solar metallicity. All those converged to
the same results obtained with the original five component base, with
the contributions from the 100 and 500 Myr populations resulting
always less than 5 \%.

\begin{figure*}
\centering
\includegraphics*[width=\columnwidth]{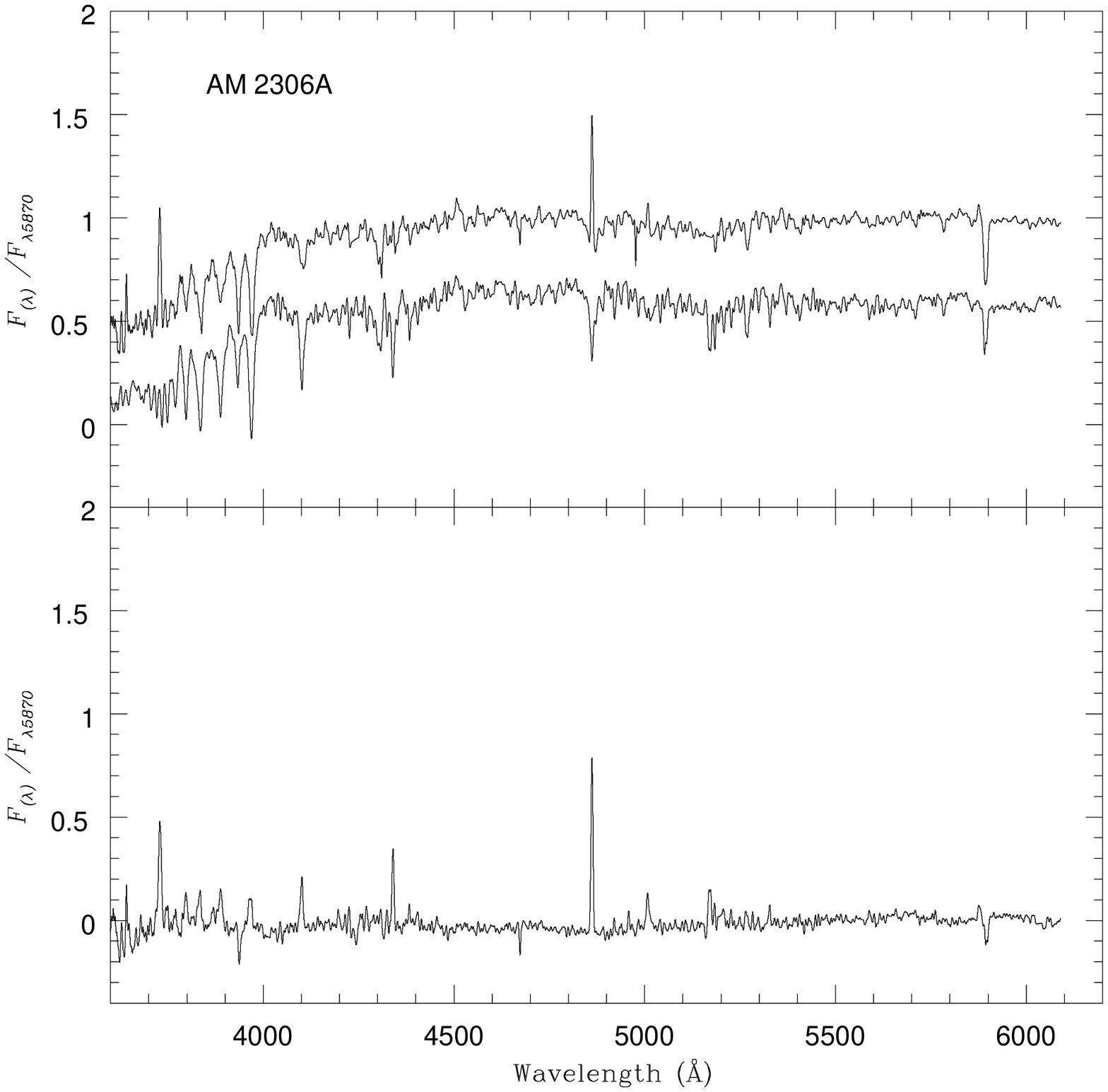}						     
\centering
\includegraphics*[width=\columnwidth]{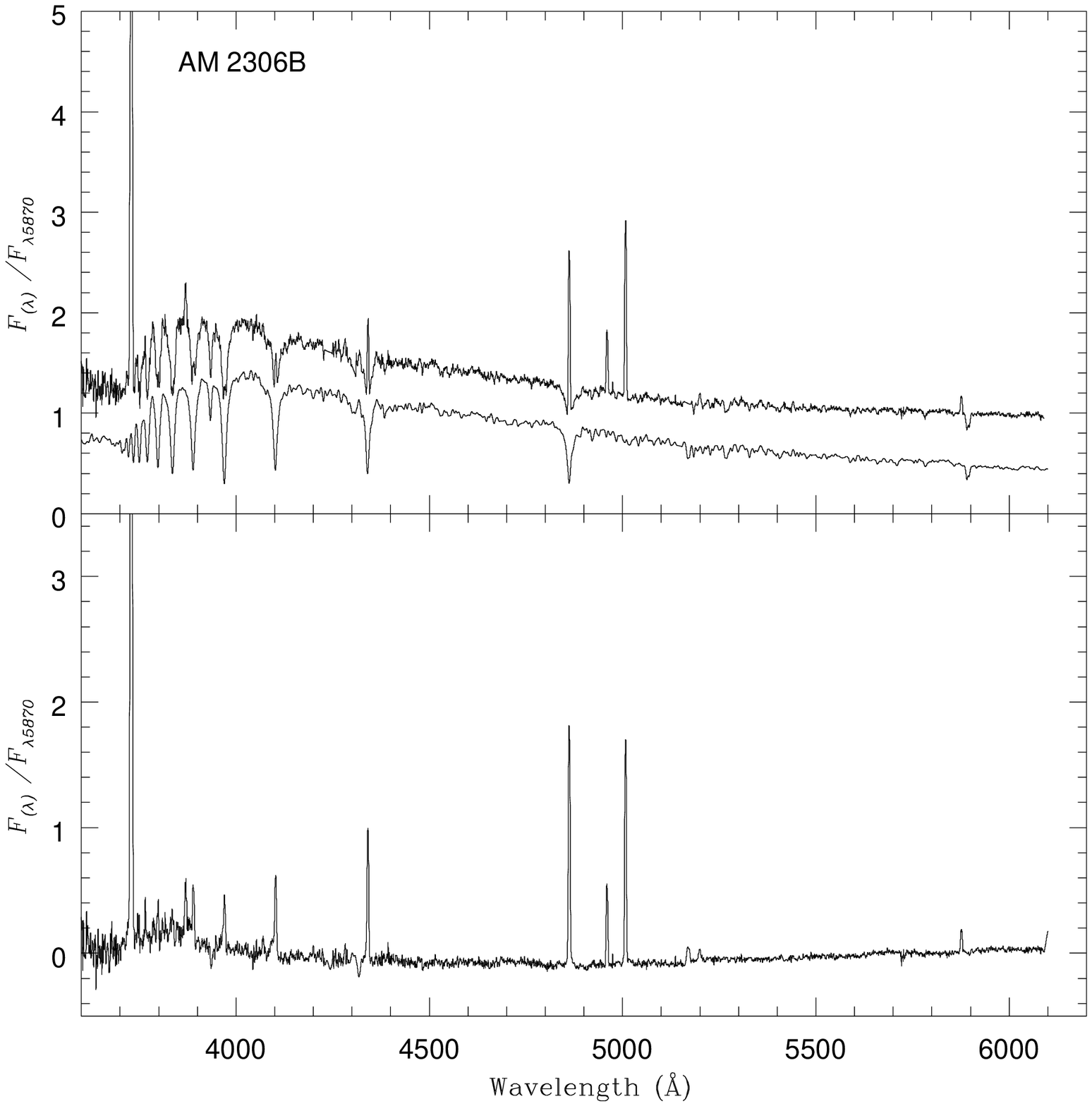}						     
\caption{Stellar population synthesis  for central bin along the major
axis of AM\,2306A (left)  and AM\,2306B (right). Top panel: spectrum
corrected for reddening  and the synthesized
spectrum (shifted by a constant). Bottom panel: pure emission spectrum.}
\label{sintesec}
\end{figure*}

Figure \ref{sintesec} show an example of the observed
spectra corrected for the reddening, the synthesized spectra and the
pure emission spectra. Table~\ref{sint_result} presents the result of
the stellar population synthesis for the individual spatial bins in each galaxy, stated as
the percentual contribution of each base element to the flux at
$\lambda\,5870\AA$. The values of $E(B-V)$ derived from the synthesis are
also given in Table~\ref{sint_result}. 

\begin{table*}
\centering
\caption{Equivalent widths of the absorption lines. The full table is available as Supplementary Material to the online version 
of this article from http://www.blackwell-synergy.com.}
\label{eqw}
\begin{tabular}{lccccccc}
\noalign{\smallskip}
\hline
\hline
\noalign{\smallskip}
&\multicolumn{7}{c}{Equivalent width (\AA)} \\
\cline{2-8}
\noalign{\smallskip}
 R(kpc)&  $\rm H11\,  \lambda\,3770$&  $\rm K CaII$&G $\lambda4300$ & \ion{Fe}{i} $\lambda\,5270$ 
 & \ion{Fe}{i} $\lambda\,5709$ &\ion{Fe}{i} $\lambda\,5782$ 
& NaD $\lambda\,5890$   \\ 
\noalign{\smallskip}
\hline
\noalign{\smallskip}
\multicolumn{8}{c}{AM\,2306A (PA=238$^{o}$)} \\
\noalign{\smallskip}
\hline
\noalign{\smallskip}
3.99 SW   &   2.85$\pm$0.40 &	   ...         &   ...	         &   ...	    &  ...        &  0.69$\pm$0.05 &    1.47$\pm$0.01  \\
2.99 SW   &   2.32$\pm$0.19 &	5.46$\pm$0.09  &   ...           &   2.35$\pm$0.03  &  ...        &  0.60$\pm$0.03 &    1.50$\pm$0.01  \\
1.99 SW   &    ...          &   6.66$\pm$0.07  &   3.80$\pm$0.20 &   2.41$\pm$0.07  &  ...	  &  0.61$\pm$0.02 &    2.15$\pm$0.02  \\
1.00 SW   &   1.12$\pm$0.12 &	5.69$\pm$0.23  &   4.00$\pm$0.31 &   2.15$\pm$0.08  &  ...        &  0.72$\pm$0.04 &	3.31$\pm$0.02  \\
0	  &    ...	    &   4.62$\pm$0.09  &   3.10$\pm$0.28 &   1.99$\pm$0.01  &  ...        &  0.75$\pm$0.02 &    4.29$\pm$0.02\\ 
\noalign{\smallskip}
\hline
\hline
\noalign{\smallskip}
\end{tabular}  
\end{table*}

\begin{figure*}
\centering
\includegraphics*[angle=-90,width=\textwidth]{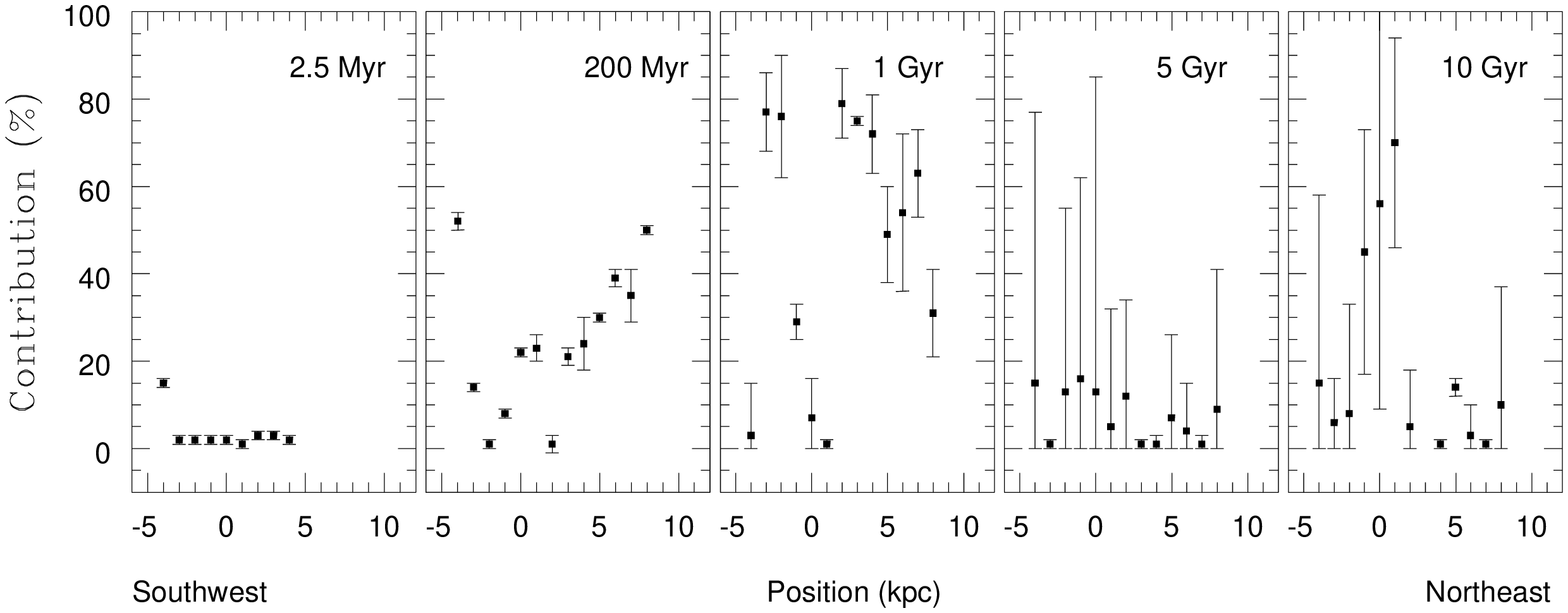}						     
\centering
\includegraphics*[angle=-90, width=\textwidth]{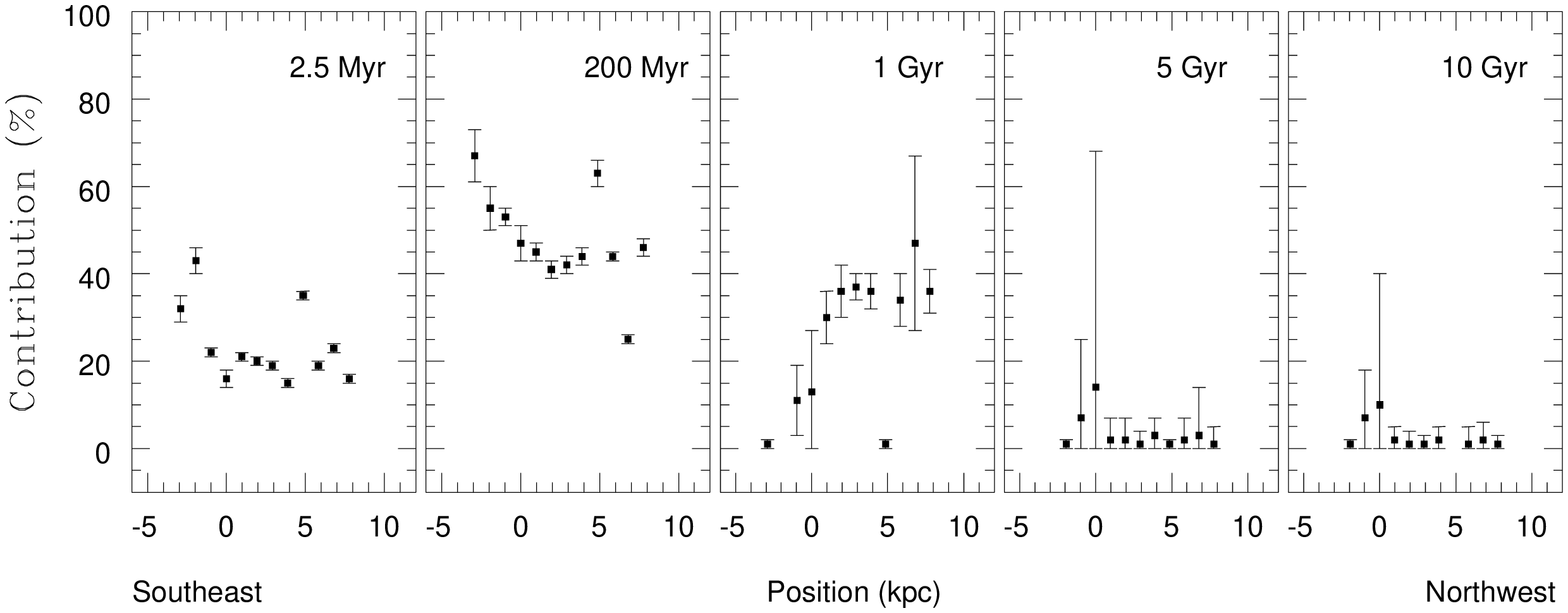}						     
\caption{Synthesis results in flux fractions as a function of distance to the center along the
major axis of  AM\,2306A (upper panel) and AM\,2306B (bottom panel).}
\label{pop_c}
\end{figure*}

The spatial variations in the contributions of the stellar population components are shown in 
Fig. \ref{pop_c}. Large variations in the contribution from the
different age components can be seen across the disk of both
galaxies. The central regions of AM\,2306A are dominated by the old
(5-10\,Gyr) population, with some significant contribution from a young
200\,Myr  and intermediate 1\,Gyr component of the spiral arms and the disk . On the
other hand, the stellar population in AM\,2306B is overall much
younger, being dominated by the 2.5\,Myr, 200\,Myr and 1\,Gyr components, which
are quite widely spread over the whole disk. 
 The star formation episode occurred about ~200\,Myr ago in both 
galaxies could be related with the perigalactic passage.

\begin{table*}
\centering
\caption{Stellar population synthesis results}
\label{sint_result}
\begin{tabular}{lrrrrrrc}
\noalign{\smallskip}
\hline
\hline
\noalign{\smallskip}
&\multicolumn{6}{c}{Flux fraction at $\lambda\,5870$ \AA} \\
\cline{2-7}
\noalign{\smallskip}
\multicolumn{1}{c}{R(kpc)}& \multicolumn{1}{c}{2.5 Myr} & \multicolumn{1}{c}{200 Myr}&
\multicolumn{1}{c}{1 Gyr} &  \multicolumn{1}{c}{5 Gyr}
  & \multicolumn{1}{c}{10 Gyr}  & \multicolumn{1}{c}{$E(B-V)$} \\ 
\noalign{\smallskip} 
\hline
\noalign{\smallskip}
\multicolumn{7}{c}{AM\,2306-721A (PA=238\degr)} \\
\noalign{\smallskip}
\hline
\noalign{\smallskip}
 3.99 SW   &	15$\pm$1    &  52$\pm$2  &  3$\pm$12   &    15$\pm$62 &   15$\pm$43 & 0.42$\pm$0.21   \\
 2.99 SW   &	 2$\pm$1    &  14$\pm$1  & 77$\pm$9    &     1$\pm$1  &    6$\pm$10 & 0.16$\pm$0.03   \\
 1.99 SW   &	 2$\pm$1    &   1$\pm$1  & 76$\pm$14   &    13$\pm$42 &    8$\pm$25 & 0.13$\pm$0.02   \\
 1.00 SW   &	 2$\pm$1    &   8$\pm$1  & 29$\pm$4    &    16$\pm$46 &   45$\pm$28 & 0.18$\pm$0.05   \\
 0	   &	 2$\pm$1    &  22$\pm$1  &  7$\pm$9    &    13$\pm$72 &   56$\pm$47 & 0.29$\pm$0.10   \\
 1.00  NE  &	 1$\pm$1    &  23$\pm$3  &  1$\pm$1    &     5$\pm$27 &   70$\pm$24 & 0.06$\pm$0.01   \\
 1.99  NE  &	 3$\pm$1    &   1$\pm$2  & 79$\pm$8    &    12$\pm$22 &    5$\pm$13 & 0.14$\pm$0.03   \\
 2.99  NE  &	 3$\pm$1    &  21$\pm$2  & 75$\pm$1    &     1$\pm$1  &    0$\pm$1  & 0.20$\pm$0.05   \\
 3.99  NE  &	 2$\pm$1    &  24$\pm$6  & 72$\pm$9    &     1$\pm$2  &    1$\pm$1  & 0.07$\pm$0.01   \\
 4.98  NE  &	 0$\pm$1    &  30$\pm$1  & 49$\pm$11   &     7$\pm$19 &   14$\pm$2  & 0.02$\pm$0.01   \\
 5.98  NE  &	 0$\pm$1    &  39$\pm$2  & 54$\pm$18   &     4$\pm$11 &    3$\pm$7  & 0.13$\pm$0.03   \\
 6.98  NE  &	 0$\pm$1    &  35$\pm$6  & 63$\pm$10   &     1$\pm$2  &    1$\pm$1  & 0.14$\pm$0.03   \\
 7.98  NE  &	 0$\pm$1    &  50$\pm$1  & 31$\pm$10   &     9$\pm$32 &   10$\pm$27 & 0.22$\pm$0.06   \\ 
\noalign{\smallskip}
\hline
\noalign{\smallskip}
\multicolumn{7}{c}{AM\,2306-721B (PA=118\degr)} \\
\noalign{\smallskip}
\hline
\noalign{\smallskip}
2.91 SE &32$\pm$3  & 67$\pm$6  &  1$\pm$1 & 0$\pm$0  	& 0$\pm$0  & 0.40$\pm$0.46	 \\
1.94 SE &43$\pm$3  & 55$\pm$5  &  0$\pm$1 & 1$\pm$1  	& 1$\pm$1  & 0.35$\pm$0.37	 \\
0.97 SE &22$\pm$1  & 53$\pm$2  & 11$\pm$8 & 7$\pm$18 	& 7$\pm$11 & 0.39$\pm$0.18	\\
0       &16$\pm$2  & 47$\pm$4  & 13$\pm$14&14$\pm$54 	&10$\pm$30 & 0.23$\pm$0.07	 \\
0.97 NW &21$\pm$1  & 45$\pm$2  & 30$\pm$6 & 2$\pm$5  	& 2$\pm$3  & 0.39$\pm$0.18	\\
1.94 NW &20$\pm$1  & 41$\pm$2  & 36$\pm$6 & 2$\pm$5  	& 1$\pm$3  & 0.31$\pm$0.12	\\
2.91 NW &19$\pm$1  & 42$\pm$2  & 37$\pm$3 & 1$\pm$3  	& 1$\pm$2  & 0.28$\pm$0.09	 \\
3.88 NW &15$\pm$1  & 44$\pm$2  & 36$\pm$4 & 3$\pm$4  	& 2$\pm$3  & 0.16$\pm$0.04	 \\
4.86 NW &35$\pm$1  & 63$\pm$3  &  1$\pm$1 & 1$\pm$1  	& 0$\pm$1  & 0.31$\pm$0.12	 \\
5.83 NW &19$\pm$1  & 44$\pm$1  & 34$\pm$6 & 2$\pm$5  	& 1$\pm$4  & 0.01$\pm$0.01	\\
6.80 NW &23$\pm$1  & 25$\pm$1  & 47$\pm$20& 3$\pm$11 	& 2$\pm$4  & 0.02$\pm$0.01	\\
7.77 NW &16$\pm$1  & 46$\pm$2  & 36$\pm$5 & 1$\pm$4  	& 1$\pm$2  & 0.01$\pm$0.01	 \\
\noalign{\smallskip}
\hline
\hline
\noalign{\smallskip}
\end{tabular}  
\end{table*}

\section{Emission line intensities and O/H abundance}
\label{emission}
Once the stellar population contribution has been determined, the
underlying absorption line spectrum can be subtracted to allow the
correct measurement and analysis of the line emission from the gaseous
component. The line intensities were measured using Gaussian line
profile fitting on the pure emission spectra. We used the {\sc IRAF splot} routine to fit
the lines, with the associated error being given as $\sigma^{2} =
\sigma_{cont}^{2} + \sigma_{line}^{2} $, where $\sigma_{cont}$ and $\sigma_{line}$ are the continuum 
rms and the Poisson error of the line flux, respectively. The residual
extinction associated with the gaseous component for each spatial bin
was calculated comparing the observed $\rm H\gamma/H\beta$ and $\rm H\alpha/H \beta$ ratios to the 
theoretical values in \citet{hummer87} for an electron temperature of 10\,000 $\mathrm{K}$ and a density of
100 $\mathrm{cm^{-3}}$. The observed emission line
intensities were then corrected by this residual extinction using the
\citet{howarth83} reddening function. 
Table \ref{lines} present the main
emission line intensities normalized to the flux in the $\rm H\beta$
line. We derived the electron densities for AM\,2306A  
from the [\ion{S}{ii}]$ \lambda\,6717/\lambda\,6731$ intensity ratios,
using the {\it temden} routine of the {\it nebular} package from 
{\it STSDAS/IRAF}, assuming an electron temperature of 10\,000 K. 
The energy levels, transition probabilities and  
collisional strength values for  [\ion{S}{ii}] were 
taken respectively from \citet{bowen60}, \citet{keenan93} and \citet{ramsbottom96}.   
 The resulting electron density obtained for AM\,2306A is in the low
density limit for the  [\ion{S}{ii}] lines, with values below $N_{{\rm
e}}= 630\, \rm cm^{-3}$, compatible with the results for both
galactic  \citep{copetti00}, and giant extragalactic \ion{H}{ii} 
regions \citep{castaneda92}. 

\begin{table*}
\label{lines}
\caption{Dereddened line fluxes $I(\lambda)$. The full table is available as Supplementary Material to the online version 
of this article from http://www.blackwell-synergy.com.}  
\begin{tabular}{rrrrrrrrrrrr}
\noalign{\smallskip}
\hline
\hline
\noalign{\smallskip}
& \multicolumn{11}{c}{AM\,2306A (PA=238\degr)} \\
\noalign{\smallskip}
\hline
\noalign{\smallskip}
\multicolumn{1}{c}{R (kpc)}& c(H$\beta$)   &\multicolumn{1}{c}{[\ion{O}{ii}] $\lambda 3727$ } 
& \multicolumn{1}{c}{\ion{H}{i} $\lambda 4340$ }& \multicolumn{1}{c}{\ion{H}{i} $\lambda 4861$ }
&\multicolumn{1}{c}{[\ion{O}{iii}] $\lambda 4949$ }  &\multicolumn{1}{c}{[\ion{O}{iii}] $\lambda 5007$ } &
\multicolumn{1}{c}{[\ion{N}{ii}] $\lambda 6548$ } & \multicolumn{1}{c}{\ion{H}{i} $\lambda 6562$ } &\multicolumn{1}{c}{[\ion{N}{ii}] $\lambda 6584$ }  &
\multicolumn{1}{c}{[\ion{S}{ii}] $\lambda 6717$ } & \multicolumn{1}{c}{[\ion{S}{ii}] $\lambda 6731$ } 
\\
\noalign{\smallskip}
\hline
\noalign{\smallskip}
12.96 SW  &    0     &     ...~~~~    &     ...~~~~ &100$\pm$3   &  40$\pm$2   &    118$\pm$4	  &  22$\pm$4   & 286$\pm$19 &   60$\pm$5   &  73$\pm$10 & 73$\pm$10  \\
11.96 SW  &    0.10  &     ...~~~~    &   47$\pm$9  &100$\pm$4   &...~~~~      &     72$\pm$4	  &  29$\pm$7	& 286$\pm$25 &   94$\pm$10  &  94$\pm$14 & 86$\pm$14  \\
10.97 SW  &    0     &     ...~~~~    &     ...~~~~ &100$\pm$5   &...~~~~      &     62$\pm$4	  &  54$\pm$13  & 286$\pm$33 &  117$\pm$17  &	...~~~~  &  ...~~~~  \\
9.97  SW  &    0     &     ...~~~~    &     ...~~~~ &100$\pm$4   &...~~~~      &     55$\pm$3	  &   ...~~~~	& 286$\pm$27 &  121$\pm$15  &	...~~~~  &  ...~~~~   \\
8.97  SW  &    0     &     ...~~~~    &     ...~~~~ &100$\pm$3   &...~~~~      &     61$\pm$3	  &  34$\pm$5	& 286$\pm$18 &  101$\pm$8   &  89$\pm$4  & 65$\pm$4   \\
\noalign{\smallskip}			       
\hline					        
\hline
\noalign{\smallskip}
\multicolumn{8}{l}{{\bf Notes:} all fluxes normalized to $F(H\beta)$ as listed in Column 5.}
\end{tabular}
\end{table*}

The spatial metallicity variations across the disk of spiral galaxies
is a fundamental parameter for understanding galaxy evolution. In
isolated spiral galaxies, it is often found that the metallicity in
the ionized interstellar medium decreases outwards
\citep{vila-costas92}, while a weaker or absent metallicity  gradient
in interacting galaxies of similar morphological type have been
pointed out as evidence of radial gas inflows  driven by tidal torques
generated during the interaction \citep{henry99,roy97}.

Determining accurate element abundances from optical spectra is
critically dependent on measuring temperature sensitive line ratios, such as
[\ion{O}{iii}]$(\lambda\,4959+\lambda\,5007)/\lambda\,4363$.
However, when doing spectroscopy of \ion{H}{ii} regions with high
metallicity and/or low excitation, temperature sensitive lines such as
[\ion{O}{iii}]$\lambda\,4363$ are found to be weak or unobservable,
and empirical indicators based on more easily measured  line ratios
have to be used to estimate metal abundances. The line ratio
$R_{23}$=([\ion{O}{ii}]$\lambda\,3727+$[\ion{O}{iii}] $\lambda\,4959+$[\ion{O}{iii}]  $\lambda\,5007)/$H$\beta$,
introduced by \citet{pagel79} has been widely used for determination of oxygen abundance 
in \ion{H}{ii} regions \citep{vila-costas92,zaritsky94,gil07}.

Although the $R_{23}$ indicator is sensitive to abundance variations,
it is double valued, presenting high to low metallicity turnover in
the range of $8< 12+\ohlog <8.3$ \citep{pilyugin05}. Another critical issue regarding the use of the
$R_{23}$ is the fact that it is also sensitive to the  ionization
level of the emission gas, especially at low metallicities. These
problems can be solved, by choosing either the low or high-metallicity
branch of  $R_{23}$, and combining it with another line indicator sensitive
to the ionization level, such as the ratio [\ion{O}{ii}]($\lambda\,3727+\lambda\,3729)/$[\ion{O}{iii}]$\lambda\,5007$
\citep{mcgaugh91}.

\begin{figure}
\centering
\includegraphics*[angle=-90,width=\columnwidth]{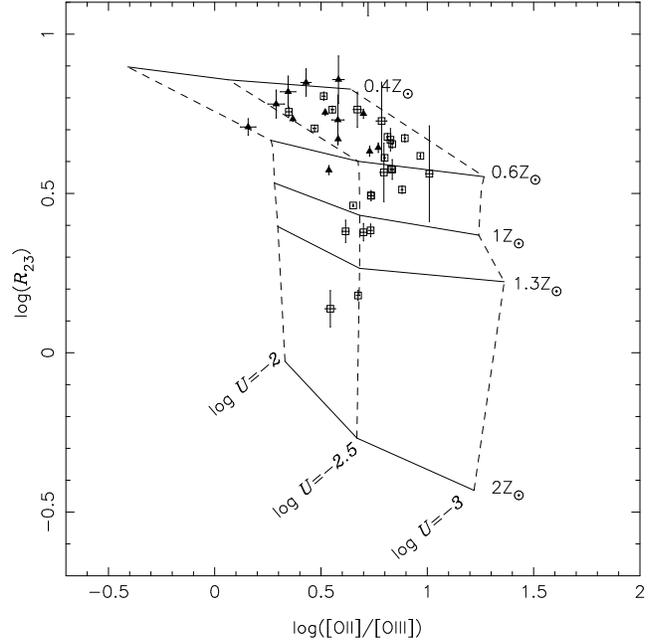}						     
\caption{The relation $\log(R_{23})$ vs. $\log$([\ion{O}{ii}]/[\ion{O}{iii}]) for the individual spatial bins in
AM\,2306A (squares) and AM\,2306B (triangles). The curves represent
the photoionization models described in the text (dashed lines
correspond to different values of the ionization parameter, solid
lines to different gas metallicity).}
\label{grade}
\end{figure}

\begin{figure}
\centering
\includegraphics*[width=\columnwidth]{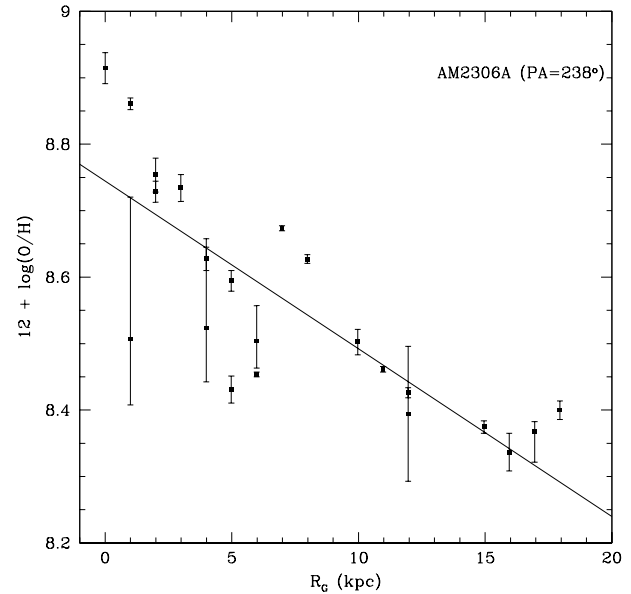}						     
\caption{The O/H abundance distribution as a function of galactocentric 
radius for the main galaxy along of the major axis. The solid line represents the fit of the data.}
\label{abund_c}
\end{figure}

\begin{figure}
\centering
\includegraphics*[width=\columnwidth]{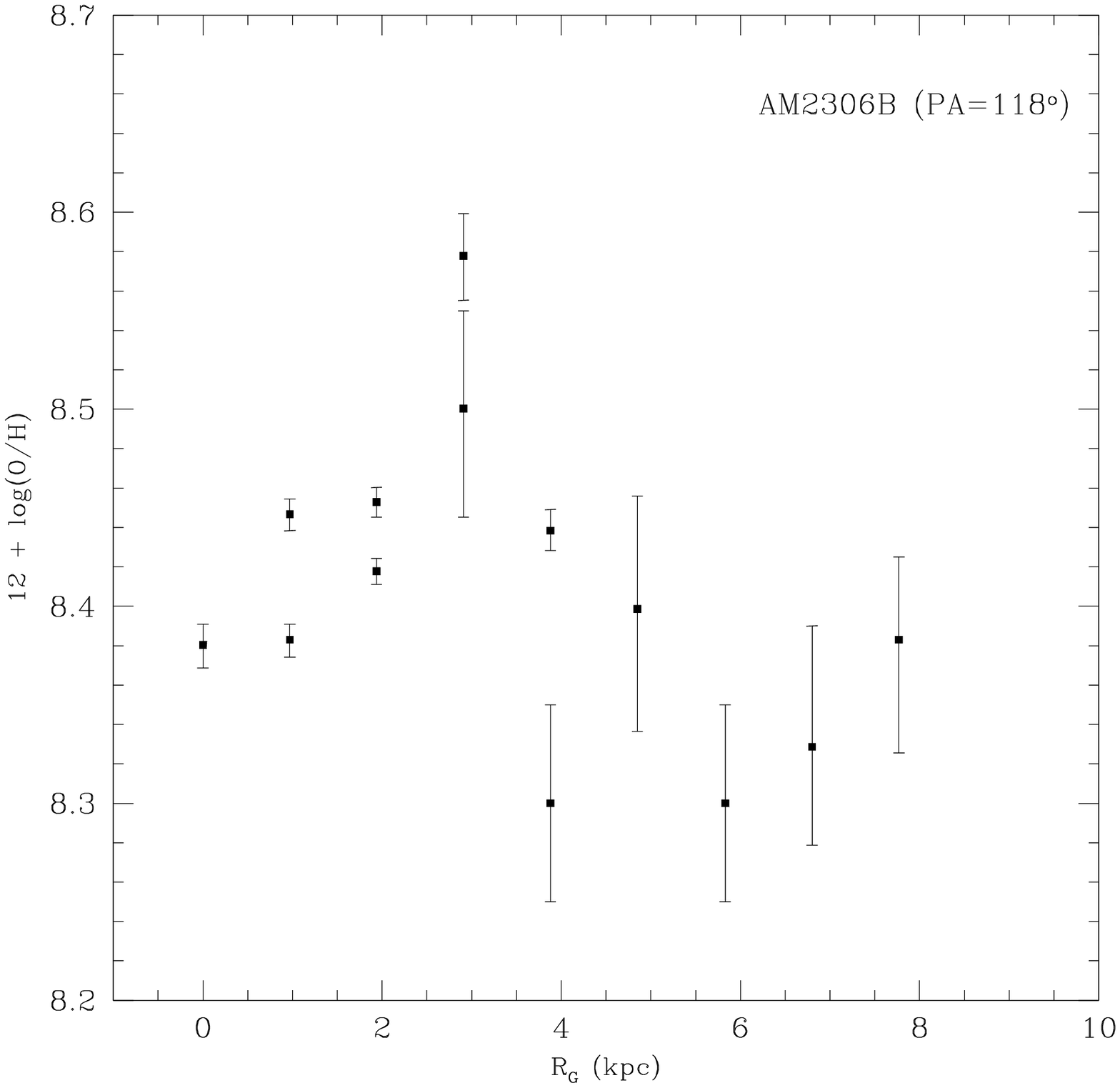}						     
\caption{The O/H abundance distribution as a function of galactocentric 
radius for the secondary galaxy along of the major axis.}
\label{abund_b}
\end{figure}

Since we do not detect any temperature-sensitive emission lines in our spectra, 
we used the $R_{23}$ indicator to estimate the metallicity comparing the observed values with  a grid of
photoionization models obtained from  the code Cloudy/96.03
\citep{ferland02}. The grid was built following the same
procedures as \cite{dors06}, with metallicities of $Z=$ 2.0,
1.1, 1.0, 0.6, and 0.4 $Z_{\odot}$, and ionization parameter
$\log\,U=$ $-2$, $-2.5$, and $-3$. The solar value of $12+\ohlog =8.69$ 
is taken from \citet{allendre01}. For each
model, the ionizing source was assumed to be a stellar cluster with
energy distribution obtained using the stellar population synthesis
code $STARBURST99$ \citep{leitherer99}, with an upper mass
limit of 100 $M_{\odot}$ and an age of 2.5 Myr.  Figure \ref{grade} shows the
$R_{23}$ versus [\ion{O}{ii}]/[\ion{O}{iii}]
diagramme, with the observed values superposed in the computed models. Open squares correspond
to regions in AM\,2306A, filled triangles to regions in AM\,2306B. Most of
the points indicate gas with solar or less than solar abundance,
with the exception of two spatial bins in AM\,2306A, corresponding
to the nuclear and the 1.0\,kpc NE regions. Figures \ref{abund_c} and \ref{abund_b} shows the estimated O/H
abundance distribution as a function of the  galactocentric distance,
$R_{G}$, for AM\,2306A and AM\,2306B, respectively. 

The disk of the main galaxy (Fig.~ \ref{abund_c}) shows a clear radial
oxygen abundance gradient, that can be fitted as a linear function
$12+\ohlog =8.75(\pm0.06)-0.025(\pm 0.007)\,R_{G}$, where $R_{G}$ is
in kpc. This abundance gradient is typical of spiral galaxies, as
found in  M\,101 \citep{kinkel94,garnett99}, and in NGC\,1365
\citep{pilyugin04,dors05}. For AM\,2306B the oxygen abundance is relatively homogeneous across
the galaxy disk, not presenting a radial oxygen abundance
gradient. The observed values are in the range $8.30<12+\ohlog <8.58$.  

 The above results are in agreement with the luminosity-metallicity
relation for gas rich galaxies
(e.g. \citealt{zaritsky94,contini02,lamareille04,tremonti04}), in the
sense that more luminous objects are more metallic. To investigate if
the interaction affected the gas metallicity in both galaxies, we
compare the central O/H value  of AM2306A and Am2306B  with isolated and pair galaxies 
from the sample of \citet{kewley06}. 
We found  the central O/H value for AM2306A is in  agreement with 
the results obtained by \citet{kewley06} for field galaxies at same
luminosity,  indicating that the interaction has not significantly affected the gas abundance in  AM2306A. 
On the other hand, the central O/H value obtained for AM2306B is 
about 0.1-0.2 dex lower than those estimated  by \citet{kewley06} for field galaxies at same luminosity, but
is in good agreement with the values obtained for galaxies in pairs.

Interestingly,  the mean
value for AM\,2306B, $12+\ohlog =8.39$, is similar to the values found
in the outer parts of AM\,2306A. If we assume that in AM\,2306B there was an oxygen abundance gradient
before the encounter with AM\,2306A, and that this gradient was destroyed by gas flows 
from the outer parts to the center of the galaxy, mixing and homogenizing the chemical composition of the
interstellar medium \citep{kewley06}, we can estimate the fraction of
infalling gas that was required to produce the metallicities that are
now observed  in AM\,2306B. For this,  we assume that, before the
interaction, AM\,2306B presented the same O/H gradient observed now in  AM\,2306A.
To reduce the initial abundance of $12+\ohlog \sim 8.75$ in the central region  to  the
current mean value of 8.39, it would be necessary that the central gas
was diluted so as to contain a fraction of 56\% of poor gas from the
outer disk plus 44\% of rich gas from the central region of the galaxy.

Another possible explanation to mix the interstellar medium and
flatten the radial gradient would be the presence of a bar
\citet{friedli94}. However, the very disturbed luminosity profile of this galaxy in
$B$, $V$ and $I$ images (FP04) do not show any evidence of such feature.

We can estimate the amount of ionized gas associated with the central
star-forming complex in AM\,2306B from the H$\alpha$ luminosity of
this region as  given by \citet{testor01}. For $R_{G}<0.2$kpc, we have
$L(H\alpha)=2.71\,\times\,10^{41}$\,erg\,s$^{-1}$  (FP04),  and
assuming a a filling factor of $f=0.30$, we find  a mass of ionized gas of 
$M_{H\,II}=8.3\times10^{5} M_{\sun}$. 
According to the above proposed scenario, 56\% of this total mass ($4.6\times10^{5} M_{\sun}$)
would be infalling gas from the outer parts of the galaxy. Since we do not see any
evidence of ionized gas flows in the radial velocities, we have to
assume that the process of gas infall has already stopped. Considering
the result of the numerical simulations, we take  a timescale of
$\sim$100\,Myr for the infall process (the gas motions were triggered by the
perigalactic passage ~250Myr ago, but are no longer observable),
resulting in an infall rate of $0.01 M_{\sun}$/yr.
This infall rate is considerably smaller than 
the  average gas inflow rate of $7 M_{\sun}$/yr to the central 1-2 kpc predicted by recent merger models
\citep{iono04}.

\section{Conclusions}
\label{final}
An observational study about the  effects of the interactions in the kinematics, stellar population and
abundances of the galaxy pair AM\,2306-721 is performed. 
The data consist of long-slit spectra in the  wavelength range of 3\,350 to 7\,130\AA\, obtained with  
the  Gemini Multi-Object Spectrograph at Gemini South. The main findings are the following:
\begin{enumerate}
\item
Rotation curves of the main and
companion galaxies with an deprojected velocity amplitude of  175 km s$^{-1}$ and
185 km s$^{-1}$ respectively were obtained.
An estimate of the dynamical mass was derived for each galaxy, 
using the deprojected velocity amplitude.
For the main galaxy, its dynamical mass is $ 1.29 \times 10^{11} M_{\sun}$
within  a radius of 18 kpc; and  for the companion galaxy, the 
estimated dynamical mass  is $ M(R)= 8.56 \times 10^{10} M_{\sun}$
 within a radius of  10.7 kpc. 
\item
In the main galaxy, radial velocity deviations from the disk rotation
of about 100 km\,s$^{-1}$ were detected, which are  probably due to  
the interaction with the companion galaxy.
\item
In order to reconstruct the history of the AM\,2306-721 system and to
predict the evolution of the encounter, we modeled the interaction
between AM\,2306A and AM\,2306B through numerical N-body/hydrodynamical simulations. The 
orbit that best reproduces the observational properties is found
to be hyperbolic, with an eccentricity $e=1.15$ and perigalacticum of
$q=5.25$ kpc; the current stage of the system would be about 250 Myr after perigalacticum. 
\item 
The spatial variations  of the stellar population components
of the galaxies were analysed  by fitting combinations of stellar population models of different ages 
(2.5 Myr, 200 Myr, 1 Gyr, 5 Gyr and 10 Gyr)  and solar
metallicity. The central regions of the main galaxy  are dominated by the old
(5-10\,Gyr) population, with some significant contribution from a young
200\,Myr  and intermediate 1\,Gyr component along the disk of the galaxy. On the
other hand, the stellar population in the companion galaxy is overall much
younger, being dominated by the 2.5\,Myr, 200\,Myr and 1\,Gyr components, which
are quite widely spread over the whole disk. 
\item
Oxygen abundance spatial profiles were obtained using a grid of
photoionization models and the $R_{23}$=([\ion{O}{ii}]$\lambda\,3727+$[\ion{O}{iii}] $\lambda\,4959+$[\ion{O}{iii}]  $\lambda\,5007)/$H$\beta$
line  ratio. The disk of the main galaxy shows a clear radial
oxygen abundance gradient,  that can be fitted as a linear function
$12+\ohlog =8.75(\pm0.06)-0.025(\pm 0.007)R_{G}$;
in the companion galaxy the oxygen abundance is relatively homogeneous across
the galaxy disk, with the observed  mean value of $12+\ohlog =8.39$ 
\item
The absence of an abundance gradient in the secondary galaxy is
 interpreted as it having been  destroyed  by gas flows from the outer
 parts to the center of the galaxy.
\end{enumerate}

\thanks{Acknowledgments}  

Based on observations obtained at the Gemini Observatory, which is operated by 
the Association of Universities for Research in Astronomy, Inc., under a 
cooperative agreement with the NSF on behalf of the Gemini partnership: the 
National Science Foundation (United States), the Science and Technology Facilities 
Council (United Kingdom), the National Research Council (Canada), CONICYT (Chile), 
the Australian Research Council (Australia), 
Minist\'erio da Ciencia e Tecnologia (Brazil) and SECYT (Argentina)

We thank  O. L. Dors, R.Riffel, R.R.Riffel and C. Bonatto for helpful discussions.
The authors would like to thank Volker Springel for providing them
with GADGET-2. 
This research has been partially supported by the Brazilian institution CNPQ and PRONEX/FAPERGS(05/11476).

\section*{SUPPLEMENTARY MATERIAL}

The following supplementary material is available for this article:\\

\noindent
{\bf Table 3.} Radial velocities. \\
{\bf Table 5.} Equivalent widths of the absorption lines.\\
{\bf Table 7.} Dereddened line fluxes $I(\lambda)$.\\

\noindent
This material is available as part of the online article from:
http://www.blackwell-synergy.com/ (this link will take you to the article abstract).

\bsp

\label{lastpage}

\end{document}